\documentclass[tradiabstract,longauth]{aa}  

\usepackage{graphicx}
\usepackage{rotating}
\usepackage{numprint}
\usepackage{lscape}
\usepackage{txfonts}
\usepackage{natbib}
\usepackage{longtable}
\usepackage{multicol}
\usepackage{hyperref}
\hypersetup{breaklinks=true}
\usepackage{color}
\usepackage{gensymb}

\newcommand{\HI}{{\rm H~{\sc i}}} 
\newcommand{\CIV}{{\rm C~{\sc iv}}} 
\newcommand{\CIII}{{\rm C~{\sc iii]}}} 
\newcommand{\AlIII}{{\rm Al~{\sc iii}}}
\newcommand{\SiIV}{{\rm Si~{\sc iv}}} 
\newcommand{\MgII}{{\rm Mg~{\sc ii}}}

\newcommand{\kms}{{\rm km}\,{\rm s}^{-1}}

\newcommand{\lya}{Lyman-$\alpha$}

\def\Sec#1{Section~\ref{s:#1}}
\def\App#1{Appendix~\ref{ap:#1}}
\def\Eq#1{Eq.~\ref{eq:#1}}
\def\Fig#1{Fig.~\ref{fig:#1}}
\def\Tab#1{Table~\ref{t:#1}}
\def\ion#1#2{{\rm #1~{\sc #2}}}

\begin{document}

   \title{The Sloan Digital Sky Survey Quasar Catalog: twelfth data release}

\author{
	    Isabelle P\^aris
	    \inst{1,2}
	    \and
	    Patrick Petitjean
	    \inst{3}
	    \and
	    Nicholas P. Ross
	    \inst{4,5}
	    \and
	    Adam D. Myers
	    \inst{6,7}
	    \and
	    \'Eric Aubourg
	    \inst{8}
	    \and
	    Alina Streblyanska
	    \inst{9,10}
	    \and
	    Stephen Bailey
	    \inst{11}
	    \and
	    \'Eric Armengaud
	    \inst{12}
	    \and
	    Nathalie Palanque-Delabrouille
	    \inst{12}
	    \and
	    Christophe Y\`eche
	    \inst{12}
	    \and
	    Fred Hamann
	    \inst{13}
	    \and
	    Michael A. Strauss
	    \inst{14}
	    \and
	    Franco D. Albareti\thanks{‘la Caixa’-Severo Ochoa Scholar}
	    \inst{,15}
	    \and
	    Jo Bovy
	    \inst{16}
	    \and
	    Dmitry Bizyaev
	    \inst{17,18}
	    \and
	    W. Niel Brandt
	    \inst{19,20,21}
	    \and
	    Marcella Brusa
	    \inst{22,23}
	    \and
	    Johannes Buchner
	    \inst{24}
	    \and
	    Johan Comparat\thanks{Severo Ochoa IFT Fellow}
	    \inst{,15}
	    \and
	    Rupert A.C. Croft
	    \inst{25}
	    \and
	    Tom Dwelly 
	    \inst{24}
	    \and
	    Xiaohui Fan
	    \inst{26}
	    \and
	    Andreu Font-Ribera
	    \inst{11}
	    \and
	    Jian Ge
	    \inst{13}
	    \and
	    Antonis Georgakakis
	    \inst{24}
	    \and
	    Patrick B. Hall
	    \inst{27}
	    \and
	    Linhua Jiang
	    \inst{28}
	    \and
	    Karen Kinemuchi
	    \inst{17}
	    \and
	    Elena Malanushenko
	    \inst{17}
	    \and
	    Viktor Malanushenko
	    \inst{17}
	    \and
	    Richard G. McMahon
	    \inst{29,30}
	    \and
	    Marie-Luise Menzel
	    \inst{24}
	    \and
	    Andrea Merloni
	    \inst{24}
	    \and
	    Kirpal Nandra
	    \inst{24}
	    \and
	    Pasquier Noterdaeme
	    \inst{3}
	    \and
	    Daniel Oravetz
	    \inst{17}
	    \and
	    Kaike Pan
	    \inst{17}
	    \and
	    Matthew M. Pieri
	    \inst{2}
	    \and
	    Francisco Prada
	    \inst{15,31,32}
	    \and
	    Mara Salvato
	    \inst{24}
	    \and
	    David J. Schlegel
	    \inst{11}
	    \and
	    Donald P. Schneider
	    \inst{19,20}
	    \and
	    Audrey Simmons
	    \inst{17}
	    \and
	    Matteo Viel
	    \inst{1,33}
	    \and
	    David H. Weinberg
	    \inst{34}
	    \and
	    Liu Zhu
	    \inst{24}
          }

   \institute{
	      INAF - Osservatorio Astronomico di Trieste, Via G. B. Tiepolo 11, I-34131 Trieste, Italy
	    \and
	      Aix Marseille Université, CNRS, LAM (Laboratoire d'Astrophysique de Marseille) UMR 7326, 13388, Marseille, France\\
	      \email{isabelle.paris@lam.fr}
	    \and
	      UPMC-CNRS, UMR 7095, Institut d’Astrophysique de Paris, 75014 Paris, France
	    \and
	      Department of Physics, Drexel University, 3141 Chestnut Street, Philadelphia, PA 19104, USA
	    \and
	      Institute for Astronomy, SUPA\footnote{Scottish Universities Physics Alliance}, University of Edinburgh, Royal Observatory, Edinburgh, EH9 3HJ, United Kingdom
	    \and
	      Department of Physics and Astronomy, University of Wyoming, Laramie, WY 82071, USA
	    \and
	      Max-Planck-Institut f\"ur Astronomie, K\"onigstuhl 17, D-69117 Heidelberg, Germany
	    \and
	      APC, Astroparticule et Cosmologie, Universit\'e Paris Diderot, CNRS/IN2P3, CEA/Irfu, Observatoire de Paris, Sorbonne Paris Cit\'e, 10 rue Alice Domon \& Léonie Duquet, 75205 Paris Cedex 13, France
	    \and
	      Instituto de Astrofisica de Canarias (IAC), E-38200 La Laguna, Tenerife, Spain
	    \and
	      Universidad de La Laguna (ULL), Dept. Astrofisica, E-38206 La Laguna, Tenerife, Spain
	    \and
	      Lawrence Berkeley National Laboratory, 1 Cyclotron Road, Berkeley, CA 94720, USA
	    \and
	      CEA, Centre de Saclay, Irfu/SPP, 91191, Gif-sur-Yvette, France
	    \and
	      Department of Astronomy, University of Florida, Gainesville, FL 32611-2055, USA
	    \and
	      Princeton University Observatory, Peyton Hall, Princeton, NJ 08544, USA
	    \and
	      Instituto de F\'{\i}sica Te\'orica (UAM/CSIC), Universidad Aut\'onoma de Madrid,  Cantoblanco, E-28049 Madrid, Spain
	    \and
	      Department of Astronomy and Astrophysics, University of Toronto, 50 St. George Street, Toronto, ON, M5S 3H4, Canada
	    \and
	      Apache Point Observatory and New Mexico State University, P.O. Box 59, Sunspot, NM, 88349-0059, USA
	    \and
	      Sternberg Astronomical Institute, Moscow State University, Moscow
	    \and
	      Department of Astronomy \& Astrophysics, The Pennsylvania State University, University Park, PA, 16802, USA
	    \and
	      Institute for Gravitation and the Cosmos, The Pennsylvania State University,University Park, PA 16802, USA
	    \and
	      Department of Physics, 104 Davey Lab, The Pennsylvania State University, University Park, PA 16802, USA
	    \and
	      Dipartimento di Fisica e Astronomia, Universit\`a di Bologna, viale Berti Pichat 6/2, 40127 Bologna, Italy
	    \and
	      INAF - Osservatorio Astronomico di Bologna, via Ranzani 1, 40127 Bologna, Italy
	    \and
	      Max-Planck-Institut f\"ur Extraterrestrische Physik, Giessenbachstrasse 1, D-85741 Garching, Germany
	    \and
	      McWilliams Center for Cosmology, Dept. of Physics, Carnegie Mellon University, Pittsburgh PA 15213
	    \and
	      Steward Observatory, University of Arizona, Tucson, AZ 85750, USA
	    \and
	      Department of Physics and Astronomy, York University, 4700 Keele St., Toronto, ON, M3J 1P3, Canada
	    \and
	      Kavli Institute for Astronomy and Astrophysics, Peking University, Beijing 100871, China
	    \and
	      Institute of Astronomy, University of Cambridge, Madingley Road, Cambridge CB3 0HA, UK
	    \and
	      Kavli Institute for Cosmology, University of Cambridge, Madingley Road, Cambridge CB3 0HA, UK
	    \and
	      Campus of International Excellence UAM+CSIC, Cantoblanco, E-28049 Madrid, Spain
	    \and
	      Instituto de Astrof\'{\i}sica de Andaluc\'{\i}a (CSIC), Glorieta de la Astronom\'{\i}a, E-18080 Granada, Spain
	    \and
	      INFN/National Institute for Nuclear Physics, Via Valerio 2,I-34127 Trieste, Italy
	    \and
	      Department of Astronomy and CCAPP, Ohio State University, Columbus, OH 43201, USA
         }
   \date{Received \today; accepted XXX}

  \abstract
   {
   We present the Data Release 12 Quasar catalog  (DR12Q) from the Baryon Oscillation Spectroscopic Survey (BOSS) of the 
   Sloan Digital Sky Survey III.
   This catalog includes all SDSS-III/BOSS objects that were spectroscopically targeted as quasar candidates during the full survey and that are confirmed as quasars via
   visual inspection of the spectra, have luminosities $M_{\rm i}[z=2] < -20.5$ (in a $\Lambda$CDM cosmology with $H_0 = 70 \ {\rm km \ s^{-1} \ Mpc ^{-1}}$,
   $\Omega _{\rm M} = 0.3$, and $\Omega _{\rm \Lambda} = 0.7$), and either display at least one emission line with a full width at half maximum (FWHM)
   larger than $500 \ {\rm km \ s^{-1}}$ or, if not, have interesting/complex absorption features. 
   The catalog also includes previously known quasars (mostly from SDSS-I and II) that were reobserved by BOSS.
   The catalog contains \numprint{297301} quasars (\numprint{272026} are new discoveries since the beginning of SDSS-III) detected over \numprint{9376} ${\rm deg ^2}$ with robust identification and redshift 
   measured by a combination of principal component eigenspectra.
   The number of quasars with $z > 2.15$ (\numprint{184101}, of which \numprint{167742} are new discoveries) is about an order of magnitude greater than the number of $z > 2.15$ quasars known prior to BOSS.
   Redshifts and FWHMs are provided for the strongest emission lines (\ion{C}{iv}, \ion{C}{iii]}, \ion{Mg}{ii}).
   The catalog identifies \numprint{29580} broad absorption line quasars and lists their characteristics.
   For each object, the catalog presents five-band ($u$, $g$, $r$, $i$, $z$) CCD-based photometry  with typical accuracy of 0.03 mag together with some information on the optical morphology and the selection criteria.
   When available, the catalog also provides information on the optical variability of quasars using SDSS and Palomar Transient Factory multi-epoch photometry.
   The catalog also contains X-ray, ultraviolet, near-infrared, and radio emission properties of the quasars, when available, from other large-area surveys.
   The calibrated digital spectra, covering the wavelength region \numprint{3600}-\numprint{10500}\AA\ at a spectral resolution in the range \numprint{1300}$< R < $\numprint{2500},
   can be retrieved from the SDSS Catalog Archive Server.
   We also provide a supplemental list of an additional \numprint{4841} quasars that have been identified serendipitously outside of the superset 
   defined to derive the main quasar catalog.
   }

   \maketitle

%
\section{Introduction}
\label{s:intro}

Since the discovery of the first quasar by \cite{schmidt1963}, each generation of spectroscopic surveys has enlarged the 
number of known quasars by roughly an order of magnitude: the Bright Quasar Survey \citep{schmidt1983} reached the 100 discoveries milestone, followed by the Large Bright Quasar Survey \citep[LBQS; ][]{hewett1995} and 
its \numprint{1000} objects, then the $\sim$\numprint{25000} quasars from the 2dF Quasar Redshift Survey 
\citep[2QZ; ][]{croom2004}, and the Sloan Digital Sky Survey \citep[SDSS; ][]{york2000} with 
over \numprint{100000} new quasars \citep{schneider2010}.
Many other surveys have also contributed to increase the number of known quasars \citep[e.g.][]{osmer1980,boyle1988,storrie1996}.
Most of these objects have redshifts of $z < 2$. 
The main goal of the Baryon Oscillation Spectroscopic Survey \citep[BOSS; ][]{dawson2013}, the main dark time survey of the 
third stage of the SDSS \citep[SDSS-III; ][]{eisenstein2011}, is to detect the characteristic scale
imprinted by baryon acoustic oscillations (BAO) in the early universe from the spatial distribution of both luminous
red galaxies at $z \sim 0.7$ \citep{anderson2012,anderson2014,alam2016} and \ion{H}{i} absorption lines in the intergalactic medium (IGM) at $z \sim 2.5$ \citep{busca2013,slosar2013,delubac2015}.
In order to achieve this goal,  
the quasar survey of the 
BOSS  was designed to discover at least 15
quasars per square degree with $2.15 \leq z \leq 3.5$ over $\sim$\numprint{10000} ${\rm deg^{2}}$ \citep{ross2012}.
The final SDSS-I/II quasar catalog of \cite{schneider2010} contained $\sim$\numprint{19000} quasars with $z > 2.15$. 
In this paper we report on the final quasar catalog from SDSS-III, which reaches a new milestone by containing
\numprint{184101} quasars at $z \geq 2.15$, \numprint{167742} of which are new discoveries.
\\

This paper presents the final SDSS-III/BOSS quasar catalog, denoted DR12Q, that compiles all the spectroscopically confirmed quasars identified
in the course of the SDSS-III/BOSS survey and released as part of the SDSS twelfth data release \citep{DR12}.
The bulk of the quasars contained in DR12Q come from the main SDSS-III/BOSS quasar target selection \citep{ross2012}. The rest of the quasars were observed
by SDSS-III/BOSS ancillary programs \citep[\numprint{46574} quasars not targeted by the SDSS-III/BOSS main quasar survey; see ][]{dawson2013,DR10,DR12}, and the \textit{Sloan Extended Quasar, ELG and LRG Survey} (SEQUELS) that is the SDSS-IV/eBOSS pilot survey \citep[\numprint{20133} quasars not targeted by the SDSS-III/BOSS main quasar survey; ][]{dawson2015,myers2015}.
In addition, we catalog quasars that are serendipitously targeted as part of the SDSS-III/BOSS galaxy surveys.
DR12Q contains \numprint{297301} unique quasars, \numprint{184101} having $z > 2.15$, over an area of \numprint{9376} ${\rm deg^2}$.
Note that this catalog does not include {\it all} the previously known SDSS quasars but only the ones that were re-observed by the SDSS-III/BOSS 
survey. Out of \numprint{105783} quasars listed in DR7Q \citep{schneider2010}, \numprint{80367} were not re-observed as part of 
SDSS-III/BOSS.
In this paper, we describe the procedures used to compile the DR12Q catalog and changes relative to the previous SDSS-III/BOSS quasar 
catalogs \citep{paris2012,paris2014}.

We summarize the target selection and observations in \Sec{observations}. We describe the visual inspection process and give the definition of the DR12Q parent sample in \Sec{construction}.
General properties of the DR12Q sample are reviewed in \Sec{sample} and the format of the file is described in \Sec{description}.
Supplementary lists are presented in \Sec{suplist}. Finally, we conclude in \Sec{conclusion}.

In the following, we will use a ${\rm \Lambda CDM}$ cosmology with ${\rm H_0 = 70 \ km \ s^{-1} \ Mpc^{-1}}$, ${\rm \Omega _M} = 0.3$, ${\rm \Omega _{\Lambda}} = 0.7$ \citep{spergel2003}.
We define a quasar as an object with a luminosity ${\rm M_i \left[ z = 2 \right] < -20.5}$ and either displaying at least one emission line with ${\rm FWHM > 500 \ km s^{-1}}$ or, if not, 
having interesting/complex absorption features.
Indeed, a few tens of objects have very weak emission lines but the \lya\ forest is seen in their spectra \citep{Diamond-Stanic2009}, and thus we include them in the DR12Q catalog. About \numprint{200} quasars with unusual broad absorption lines are also included in our catalog \citep{hall2002} even though they do not formally meet the requirement on emission-line width.
All magnitudes quoted here are Point Spread Function (PSF) magnitudes \citep{stoughton2002} and are corrected for Galactic extinction \citep{schlegel1998}.

%
\section{Observations}
\label{s:observations}

The measurement of the clustering properties of the IGM using the \lya\ forest to constrain the BAO scale at $z \sim 2.5$
at the percent level  requires a surface density of at least  15 quasars ${\rm deg^{-2}}$
 in the redshift range 2.15--3.5 over about \numprint{10000} square degrees \citep{mcdonald2007}. 
To meet these requirements,
the five-year SDSS-III/BOSS program observed 
about \numprint{184000} quasars with $2.15 < z < 3.5$ over \numprint{9376} ${\rm deg^2}$. 
The precision reached is of order 4.5\% on the angular diameter distance and 2.6\% in the Hubble constant at $z \sim 2.5$ \citep{eisenstein2011,dawson2013,DR12}.
The BAO signal was detected in the auto-correlation of the \lya\ forest \citep{busca2013,slosar2013,delubac2015} and in the cross-correlation of quasars with the \lya\ forest
\citep{font2014} at $z \sim 2.5$ using previous data releases of SDSS-III \citep{DR9,DR10}.

	\subsection{Imaging data}

The SDSS-III/BOSS quasar target selection \citep[][ and references therein]{ross2012} is based on the SDSS-DR8 imaging data \citep{DR8}
which include the SDSS-I/II data with an extension to the South Galactic Cap (SGC).
Imaging data were gathered using a dedicated 2.5m wide-field telescope \citep{gunn2006} to collect light for a camera with 30 2k$\times$2k 
CCDs \citep{gunn1998} over five broad bands - \textit{ugriz} \citep{fukugita1996}.
A total of \numprint{14555} unique square degrees of the sky were imaged by this camera, including contiguous areas of $\sim$\numprint{7500}~${\rm deg^2}$ in the North
Galactic Cap (NGC) and $\sim$\numprint{3100} ${\rm deg^2}$ in the SGC that comprise the uniform ``Legacy'' areas of the SDSS \citep{DR8}.
These data were acquired on dark photometric nights of good seeing \citep{hogg2001}.
Objects were detected and their properties were measured by the photometric pipeline \citep{lupton2001,stoughton2002} and calibrated 
photometrically \citep{smith2002,ivezic2004,tucker2006,padmanabhan2008}, and astrometrically \citep{pier2003}.
The SDSS-III/BOSS limiting magnitudes for quasar target selection are $r \leq 21.85$ or $g \leq 22$.

	\subsection{Target selection}
	\label{s:QTS}

	    \subsubsection{SDSS-III/BOSS main survey}

The measurement of clustering in the IGM
is independent of the properties of background quasars used to trace the IGM, allowing for a heterogeneous target selection
to be adopted in order to achieve the main cosmological goals of BOSS.
Conversely there is the desire to perform demographic measurements and quasar physics studies using a uniformly-selected quasar sample
\citep[e.g. ][]{white2012,ross2013,eftekharzadeh2015}.
Thus, a strategy that mixes a heterogeneous selection to maximize the surface density of $z > 2.15$ quasars,
with a uniform subsample has been adopted by SDSS-III/BOSS \citep{ross2012}.

On average, 40 fibers per square degree were allocated to the SDSS-III/BOSS quasar survey.
About half of these fibers are selected by a single target selection in order to create a uniform (``CORE'') sample.
Based on tests performed during the first year of BOSS observations, the CORE selection was performed by the
XDQSO method \citep{bovy2011}.
The second half of the fibers are dedicated to a ``BONUS'' sample designed to maximize the $z>2.15$ quasar surface density.
The output of several different methods -- a neural network combinator \citep{yeche2010}, a Kernel Density Estimator \citep[KDE; ][]{richards2004,richards2009},
a likelihood method \citep{kirkpatrick2011}, and the XDQSO method with lower likelihood than in the CORE sample -- are combined to select
the BONUS quasar targets.
Where available, near-infrared data from UKIDSS \citep{lawrence2007}, ultraviolet data from GALEX \citep{martin2005}, along with deeper coadded
imaging in overlapping SDSS runs \citep{DR8}, were also incorporated to increase the high-redshift quasar yield \citep{Bovy2012}.
In addition, all point sources that match the FIRST radio survey \citep[July 2008 version; ][]{becker1995} within a 1\arcsec\ radius and have $(u-g)>0.4$
(to filter out $z < 2.15$ quasars) are included in the quasar target selection.

To take advantage of the improved throughput of the SDSS spectrographs, previously known quasars from the SDSS-DR7 \citep{schneider2010}, the 2QZ \citep{croom2004}, the AAT-UKIDSS-DSS (AUS), 
the MMT-BOSS pilot survey \citep[Appendix C in ][]{ross2012} and quasars observed at high spectral resolution by VLT/UVES and Keck/HIRES were re-targeted.
During the first two years of BOSS observations, known quasars with $z > 2.15$ were systematically re-observed.
We extended the re-observations to known quasars with $z>1.8$ starting from Year 3 in order to better characterize metals in the \lya\ forest.

	    \subsubsection{SDSS-III/BOSS ancillary programs}

In addition to the main survey, about 5\% of the SDSS-III/BOSS fibers are allocated to ancillary programs that cover a variety of scientific goals. 
The programs are described in the Appendix and Tables~6 and 7 of \cite{dawson2013} and \S 4.2 of \cite{DR10}.
Ancillary programs whose observation started after MJD~=~56107 (29th June 2012) are detailed  in Appendix A of \cite{DR12}.
The full list of ancillary programs targeting quasars (and their associated target selection bits) that are included in the present catalog 
is provided in \App{QTSflag}.


	    \subsubsection{The Sloan Extended Quasar, ELG, and LRG Survey (SEQUELS)}

A significant fraction of the remaining dark time available during the final year of SDSS-III was dedicated to a pilot survey to prepare for the extended Baryonic Oscillation Spectroscopic Survey \citep[eBOSS; ][]{dawson2015}, the Time 
Domain Spectroscopic Survey \citep[TDSS; ][]{morganson2015} and the SPectroscopic Identification of eROSITA Sources survey (SPIDERS): the Sloan Extended QUasar, ELG and LRG Survey (SEQUELS).
The SEQUELS targets are tracked with the {\tt EBOSS\_TARGET0} flag in DR12Q. The {\tt EBOSS\_TARGET0} flag is detailed in Appendix A.3 and Table 7 of \cite{DR12}.
We provide here a brief summary of the various target selections applied to this sample.\\

The main SEQUELS subsample is based upon the eBOSS quasar survey \citep{myers2015}:
\begin{itemize}
 \item The eBOSS ``CORE'' sample is intended to recover sufficient quasars in the redshift range $0.9 < z < 2.2$ for the cosmological goals
 of eBOSS and additional quasars at $z > 2.2$ to supplement BOSS. The CORE sample homogeneously targets quasars at all redshifts $z > 0.9$
 based on the XDQSOz method \citep{Bovy2012} in the  optical and a WISE-optical color cut.
 In order to be selected, it is required that point sources have a XDQSOz probability to be a $z > 0.9$ quasar larger than 0.2 \textit{and} pass the color cut 
 $m_{\rm opt} - m_{\rm WISE} \geq \left( g -i \right) + 3$ where $m_{\rm opt}$ is a weighted stacked magnitude in the $g$, $r$ and $i$ bands and $m_{\rm WISE}$ is a weighted stacked
 magnitude in the W1 and W2 bands.
Quasar candidates have $g < 22$ \textit{or} $r < 22$ with a surface density of confirmed new quasars (at any redshifts) of
$\sim 70 \ {\rm deg^{-2}}$.
 \item PTF sample: quasars over a wide range of redshifts are selected via their photometric variability measured from the Palomar Transient Factory \citep[PTF; ][]{rau2009,law2009}.
 Quasar targets have $r > 19$ and $g < 22.5$ and provide an additional 3-4 $z > 2.1$ quasars per ${\rm deg^2}$.
 \item Known quasars with low quality SDSS-III/BOSS spectra ($0.75 < {\rm S/N \ per \ pixel} < 3$) or with bad spectra are re-observed.
 \item uvxts==1 objects in the KDE catalog of \cite{richards2009} are targeted (this target class will be discontinued for eBOSS).
 \item Finally, quasars within within 1\arcsec of a radio detection in the FIRST point source catalog \citep{becker1995} are targeted.
\end{itemize}

TDSS target point sources 
are selected to be variable in the $g$, $r$ and $i$ bands using the SDSS-DR9 
imaging data \citep{DR9} and the multi-epoch Pan-STARRS1 (PS1) photometry
\citep{kaiser2002,kaiser2010}. 
The survey does not specifically target quasars but a significant fraction of targets belong to this class \citep{morganson2015}, and are thus included in the parent sample for the quasar catalog.

Finally, the AGN component of SPIDERS targets X-ray sources detected in the concatenation of the Bright and Faint ROSAT All Sky Survey (RASS) catalogs \citep{voges1999,voges2000} and that have an
optical counterpart detected in the DR9 imaging data \citep{DR9}. Objects with $17 < r < 22$ that lie within 1\arcmin\ of a RASS source are targeted.

	\subsection{Spectroscopy}

\begin{figure}[htbp]
	\centering{\includegraphics[angle=-90,width=\linewidth]{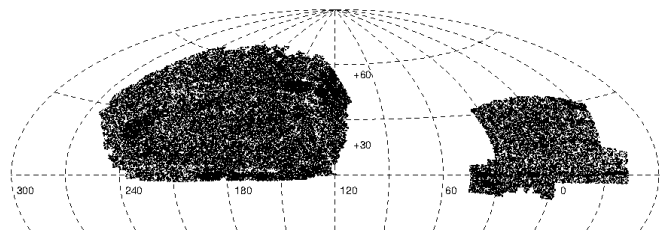}}
\caption{Distribution on the sky of the SDSS-DR12/BOSS spectroscopy in J2000 equatorial coordinates (expressed in decimal degrees).
}
\label{fig:SkyCoverage}
\end{figure}

\begin{figure}[htbp]
 \centering{\includegraphics[width=75mm]{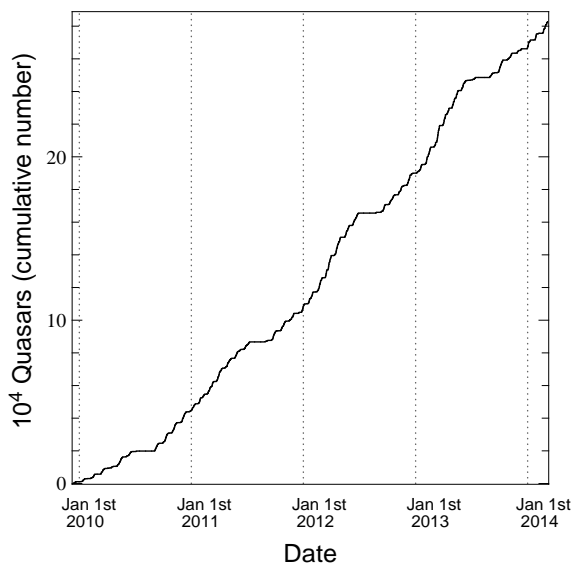}}
 \caption{Cumulative number of quasars as a function of observation date during the whole SDSS-III/BOSS survey. 
 A total of \numprint{297301} quasars was spectroscopically confirmed.
 The ``flat portions'' in this figure correspond to summer shutdowns.}
\label{fig:progress_plot}
\end{figure}

Quasars selected by the various selection algorithms are observed with the BOSS spectrographs whose resolution varies from  
$\sim$\numprint{1300} at \numprint{3600}~\AA~ to \numprint{2500} at \numprint{10000}~\AA~ \citep{smee2013}.
Spectroscopic observations are obtained in a series of at least three 15-minute exposures.
Additional exposures are taken until the squared signal-to-noise ratio per pixel, (S/N)$^2$, reaches the survey-quality threshold for each CCD.
These thresholds are ${\rm (S/N)^2} \geq 22$ at $i$-band magnitude 21 for the red camera
and ${\rm (S/N)^2} \geq 10$ at $g$-band magnitude 22 for the blue camera (Galactic extinction-corrected magnitudes). 
The spectroscopic reduction pipeline for the BOSS spectra is described  
in \cite{bolton2012}.
SDSS-III uses plates covered by \numprint{1000} fibers that can be placed such that, ideally, up to \numprint{1000} spectra
can be obtained simultaneously. The plates are tiled in a manner which allows them to overlap \citep{dawson2013}.
A total of \numprint{2438} plates were observed between December 2009 and June 2014; 
some were observed multiple times. 
In total, \numprint{297301} unique quasars have been spectroscopically confirmed based on our visual inspection described below. \Fig{SkyCoverage} shows the observed sky regions.
The total area covered by the SDSS-III/BOSS is \numprint{9376} ${\rm deg^2}$.
\Fig{progress_plot} shows the number of spectroscopically confirmed quasars with respect to their observation date.

%
\section{Construction of the DR12Q catalog}
\label{s:construction}

Spectra of all the SDSS-III/BOSS quasar candidates, all the SDSS-III quasar targets from ancillary programs and all the objects classified robustly as $z \geq 2$ quasars
by the SDSS pipeline \citep{bolton2012} among the galaxy targets \citep{reid2015} have been visually inspected.
The SDSS-DR12 quasar catalog is built upon the output of the visual inspection of \numprint{546856} such quasar candidates.

	\subsection{Visual inspection process}
	\label{s:vi_process}

The spectra of quasar candidates are reduced by the SDSS pipeline\footnote{The software used to reduce SDSS
data is called idlspec2d. Its DR12 version is v5\_7\_0 for spectra taken prior to MJD~=~56755 and v5\_7\_3 after that date. Details can be found at http://www.sdss3.org/dr12/software and in
\cite{bolton2012}.}, which provides a classification ({\tt QSO}, {\tt STAR} or {\tt GALAXY}) and a redshift.
To achieve this, a library of stellar templates, and a principal component analysis (PCA) decomposition of galaxy and quasar spectra are fitted to each spectrum.
Each class of templates is fitted in a given range of redshift: 
galaxies from $z = -0.01$ to $1.00$, quasars from $z = 0.0033$ to $7.00$, and 
stars from $z = -0.004$ to $0.004$ ($\pm$\numprint{1200}$\ {\rm km \ s^{-1}}$).

The pipeline thus provides the overall best fit (in terms of the lowest reduced $\chi ^2$)  and a redshift
for each template and thus a classification for the object.
A quality flag ({\tt ZWARNING}) is associated with each measured redshift. 
{\tt ZWARNING}~=~0 indicates that the pipeline has high confidence in its estimate.
This output is used as the starting point for the visual inspection process \citep[see ][ for details]{bolton2012}.

\begin{figure*}[htbp]
 \centering{\includegraphics[angle=-90,width=150mm]{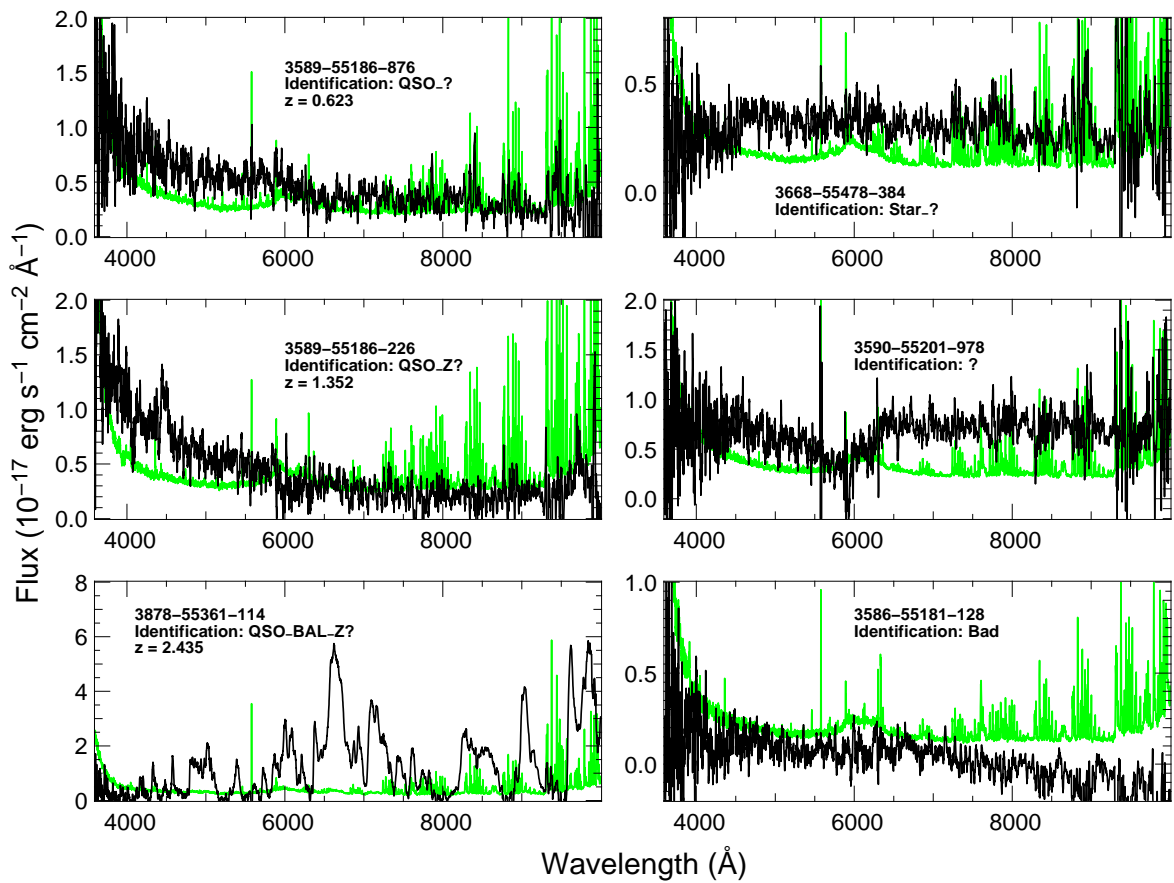}}
 \caption{
 Examples of spectra identified as {\tt QSO\_?} (top left), {\tt QSO\_Z?} (middle left), {\tt QSO\_BAL\_Z?} (bottom left), {\tt Star\_?} (top right), {\tt ?} (middle right), and {\tt Bad} (bottom right).
 The fluxes are shown in black and associated errors in green.
  All the spectra are smoothed over 10 pixels.
 }
 \label{fig:exUncertain}
\end{figure*}

Based on the initial SDSS pipeline classification and redshift, the quasar candidates are split into four categories:
low-z ($z<2$) and high-z ($z\geq2$) quasars, stars and ``others''.
The spectra are visually inspected through a dedicated webpage on which all spectra are matched to targeting information, imaging data and SDSS-I/II spectroscopy.
The inspection is performed plate by plate for each category described above with a list of spectra whose identification and redshift can be validated at once. If an object requires a more thorough inspection, it can be inspected on a single page on which it is possible to change the redshift and identification of the object.
Whenever a change of identification and/or redshift is necessary, it is done \textit{by eye}.
Additional flags are also set during the visual inspection when a Damped \lya\ system or broad absorption lines (see \Sec{bal} for further details) are seen. Finally, quality flags are set if there is an obvious reduction problem.
When the redshift and identification are unambiguous based on this inspection, each object is classified {\tt QSO}, 
{\tt Star} or {\tt Galaxy}
and a redshift is adopted.
If the pipeline redshift does not need to be modified, the SDSS pipeline redshift ({\tt Z\_PIPE}) and the visual inspection redshift ({\tt Z\_VI}) are identical. If not, the redshift is set by eye with an accuracy that cannot be better than $\Delta z \sim 0.003$.
In some cases, objects undoubtedly are quasars but their redshift is uncertain; these objects are classified
as {\tt QSO\_Z?} or {\tt QSO\_BAL\_Z?} 
when the spectrum exhibits strong broad absorption lines.
When the object identification is unsure 
they are classified either as {\tt QSO\_?} or as {\tt Star\_?}.
For spectra with a moderate S/N but with a highly uncertain identification, the object is classified as {\tt ?}.
Finally, when the S/N is too low or the spectrum is affected by poor sky subtraction or has portions missing
needed that prevent an identification, it is classified as {\tt Bad}. The
distinction between {\tt Bad} and {\tt ?} becomes somewhat subjective as the S/N decreases. 
Examples of spectra to illustrate these categories are shown in \Fig{exUncertain}.

Most of the quasars are observed as part of the main SDSS-III/BOSS quasar survey whose target selection is
described in \cite{ross2012} or ancillary programs targeting quasars.
In addition, 
we also included serendipitous quasars that were found among galaxy targets.
Because these latter spectra have usually low S/N, their exact identification is more difficult. 
Thus, specifically for these objects we adopted a
binary visual inspection classification: either an object is a {\tt QSO} or  is {\tt not\_a\_QSO}.

	\subsection{Definition of the DR12Q parent sample}
	\label{s:defDR12}

We selected objects for visual inspection from among the following targets:
\begin{itemize}
 \item quasar targets from the main SDSS-III/BOSS quasar survey: these targets can be identified with the {\tt BOSS\_TARGET1}
 target selection flag \citep[See ][for more details]{ross2012};
 \item ancillary programs that target quasars: these can be identified with the {\tt ANCILLARY\_TARGET1} and {\tt ANCILLARY\_TARGET2}
 target selection flags that are fully described in Appendix B of \cite{dawson2013} and Appendix A of \cite{DR12};
 \item quasar targets from the SDSS-IV pilot survey that can be identified with the {\tt EBOSS\_TARGET0} flag. The meaning of each
 value of this target selection bit is fully described in Appendix A of \cite{DR12};
 \item SDSS-III/BOSS galaxy targets that are identified by the pipeline as either {\tt QSO} with ${\rm z} > 2$ and 
{\tt ZWARNING} set to 0, or {\tt GALAXY} with the subclass {\tt BROADLINE}.
\end{itemize}
These criteria led to a superset containing \numprint{546856} objects among which we identified
\numprint{297301} {\tt QSO}, \numprint{1623} {\tt QSO\_?}, \numprint{1490} {\tt QSO\_Z?}, \numprint{22804} {\tt Galaxy},
\numprint{205282}  {\tt Star}, \numprint{2640} {\tt Star\_?}, \numprint{6292} {\tt Bad}, and \numprint{2386} {\tt ?}.
Among the galaxy targets, \numprint{6844} objects were identified as {\tt not a QSO} after their visual inspection. 
Finally, a total of \numprint{192} quasar targets could not be visually inspected due to bad photometric information.
The identification or redshift of \numprint{53413} objects (out of \numprint{546856}) changed with respect to the SDSS
pipeline \citep{bolton2012}. A total of 83\% of objects classified by the SDSS pipeline have {\tt ZWARNING}~=~0, i.e. their identification
and redshift are considered secure. Among this subset, the identification or redshift of 3.3\% of the objects changed after
visual inspection.
In such a case, the redshift is adjusted \textit{by eye}.
For the remaining 96.7\%, we kept the SDSS pipeline redshift (and associated error) and identification.
Emission-line confusions are the main source of catastrophic redshift failures, and wrong identifications mostly affect stars
that are confused with $z \sim 1$ quasars because of poor sky subtraction at the red end of their spectra.
The remaining 17\% of the overall sample have {\tt ZWARNING}$>$0, i.e. the pipeline output may not be reliable. For
these objects, 41.1\% of the identification or redshift changed after inspection, following the same procedure as previously.
The result of the visual inspection is summarized in \Tab{result_vi}.
The SDSS-DR12 quasar catalog lists all the firmly confirmed quasars, i.e. those identified as {\tt QSO} and {\tt QSO\_BAL}.

\begin{table*}
\centering
\begin{tabular}{l r r r}
\hline
\hline
Classification & \# pipeline & \# pipeline & \# visual \\ 
  &   & with {\tt ZWARNING}=0 & inspection \\ 
\hline
\hline
{\tt QSO} & \numprint{328433} & \numprint{287486} & \numprint{297301} \\ 
{\tt QSO} with $z>2.15$ & (\numprint{193064}) & (\numprint{179691}) & (\numprint{184101}) \\ 
{\tt QSO\_?} & - & - & \numprint{1623} \\ 
{\tt QSO\_Z?} & - & - & \numprint{1490} \\ 
{\tt Galaxy} & \numprint{41113} & \numprint{30373} & \numprint{22804} \\ 
{\tt Star} & \numprint{177310} & \numprint{135722} & \numprint{205282} \\ 
{\tt Star\_?} & - & - & \numprint{2640} \\ 
{\tt Bad} & - & - & \numprint{6292} \\ 
{\tt ?} & - & - & \numprint{2386} \\ 
Missing & - & - & \numprint{194} \\ 
Not\_a\_QSO (galaxy targets) & - & - & \numprint{6844} \\ 
\hline
 Total & \numprint{546856} & \numprint{453581} & \numprint{546856} \\ 
\hline
\multicolumn{4}{l}{$^a$ Quasars not visually inspected because of bad photometric information.}
\end{tabular}
\caption{ Number of objects identified as such by the pipeline with any {\tt ZWARNING} value (second column) and with {\tt ZWARNING}~=~0 (third column), and after
the visual inspection (fourth column).}
\label{t:result_vi}
\end{table*}

Together with the quasar catalog, the full result of the visual inspection, i.e. the identification of all
visually inspected objects, is made available as part of the superset file from which we derive this
catalog. 
The identification of each object described in \Sec{vi_process} (e.g. {\tt QSO}, {\tt QSO\_Z?}, {\tt Bad}) is encoded with two parameters in the superset file: {\tt CLASS\_PERSON} and {\tt Z\_CONF\_PERSON}. The former refers to the classification itself,  {\tt QSO}, {\tt Star} or {\tt Galaxy}, while the latter refers to our level of confidence for the redshift we associate to each object.
The relation between these two parameters and the identification from \Sec{vi_process} is given in \Tab{VI_PIPE}.
We refer the reader to \App{Superset_format} for the detailed format of the DR12 superset file, which is
provided in FITS format \citep{wells1981}.

%
\begin{table*}
\centering                        
\begin{tabular}{c c c c c}        
\hline\hline
                & \multicolumn{4}{c}{z\_conf\_person}  \\                                        
                & 0                  & 1                  & 2                  & 3        \\
class\_person   &                    &                    &                    &          \\
\hline
0               & Not inspected      & {\tt ?}            & -                  & -        \\
1               & -                  & {\tt Star\_?}      & -                  & {\tt Star}     \\
2               & Not a quasar $^a$  & -                  & -                  & -              \\
3               & -                  & {\tt QSO\_?}       & {\tt QSO\_Z?}      & {\tt QSO}      \\
4               & -                  & -                  & -                  & {\tt Galaxy}   \\
30              & -                  & -                  & {\tt QSO\_BAL\_Z?} & {\tt QSO\_BAL} \\
\hline   
\multicolumn{5}{l}{$^a$ Galaxy targets}                         
\end{tabular}
\caption{The visual inspection classification, corresponding to the combination of the {\tt class\_person} (first column) and {\tt z\_conf\_person} (first row) values 
provided in the superset file described in \App{Superset_format}. 
}
\label{t:VI_PIPE}
\end{table*}

%
\section{Summary of the sample}
\label{s:sample}

    \subsection{Broad view}

The DR12Q catalog contains a total of \numprint{297301} unique quasars, among which \numprint{184101}
have $z > 2.15$. This represents an increase of 78\% compared to the full SDSS-DR10 quasar catalog \citep{paris2014},
and \numprint{121923} quasars are new discoveries compared to the previous release.
The full SDSS-III spectroscopic survey covers a total footprint of \numprint{9376} ${\rm deg^2}$,
leading to an average surface density of $z>2.15$ quasars of $19.64 \ {\rm deg^{-2}}$.
A total of \numprint{222329} quasars, \numprint{167007} having $z > 2.15$, are located in the
SDSS-III/BOSS main quasar survey and are targeted as described in \cite{ross2012}.\footnote{Quasars that have bits 10, 11, 40 or 41 set for {\tt BOSS\_TARGET1}}
Additional BOSS-related programs\footnote{Quasars that do not have bits 10, 11, 40 or 41 set for {\tt BOSS\_TARGET1} and that have {\tt EBOSS\_TARGET0}~==~0.}
targeted \numprint{54639} quasars, and SEQUELS\footnote{Quasars that have {\tt EBOSS\_TARGET0}$>$0} targeted a total of \numprint{20333} confirmed quasars.

\begin{figure*}[htbp]
	\centering{
		\begin{minipage}{.45\linewidth}
			\centering{\includegraphics[width=75mm]{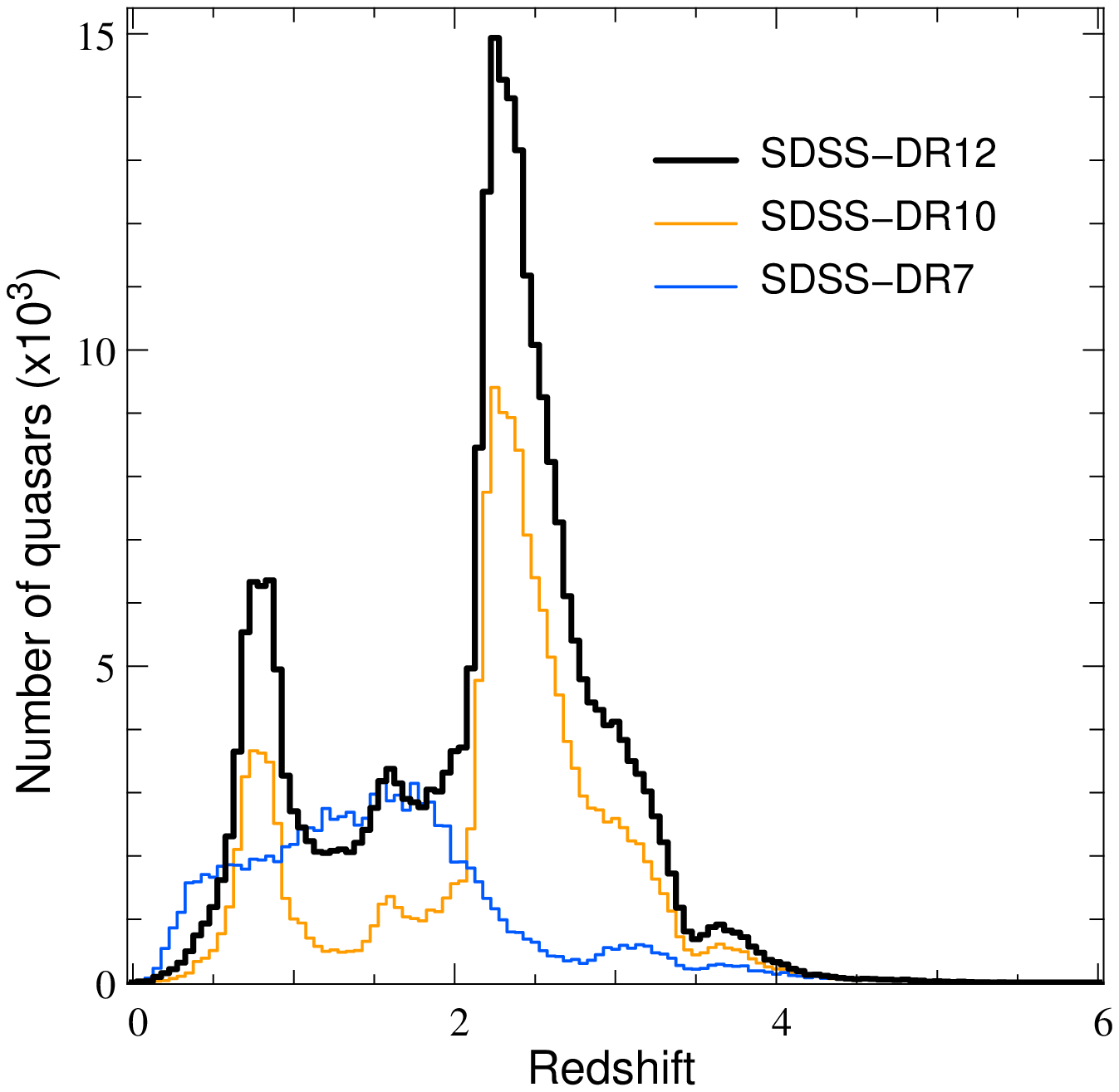}}
		\end{minipage}
		\hfill
		\begin{minipage}{.45\linewidth}
			\centering{\includegraphics[width=75mm]{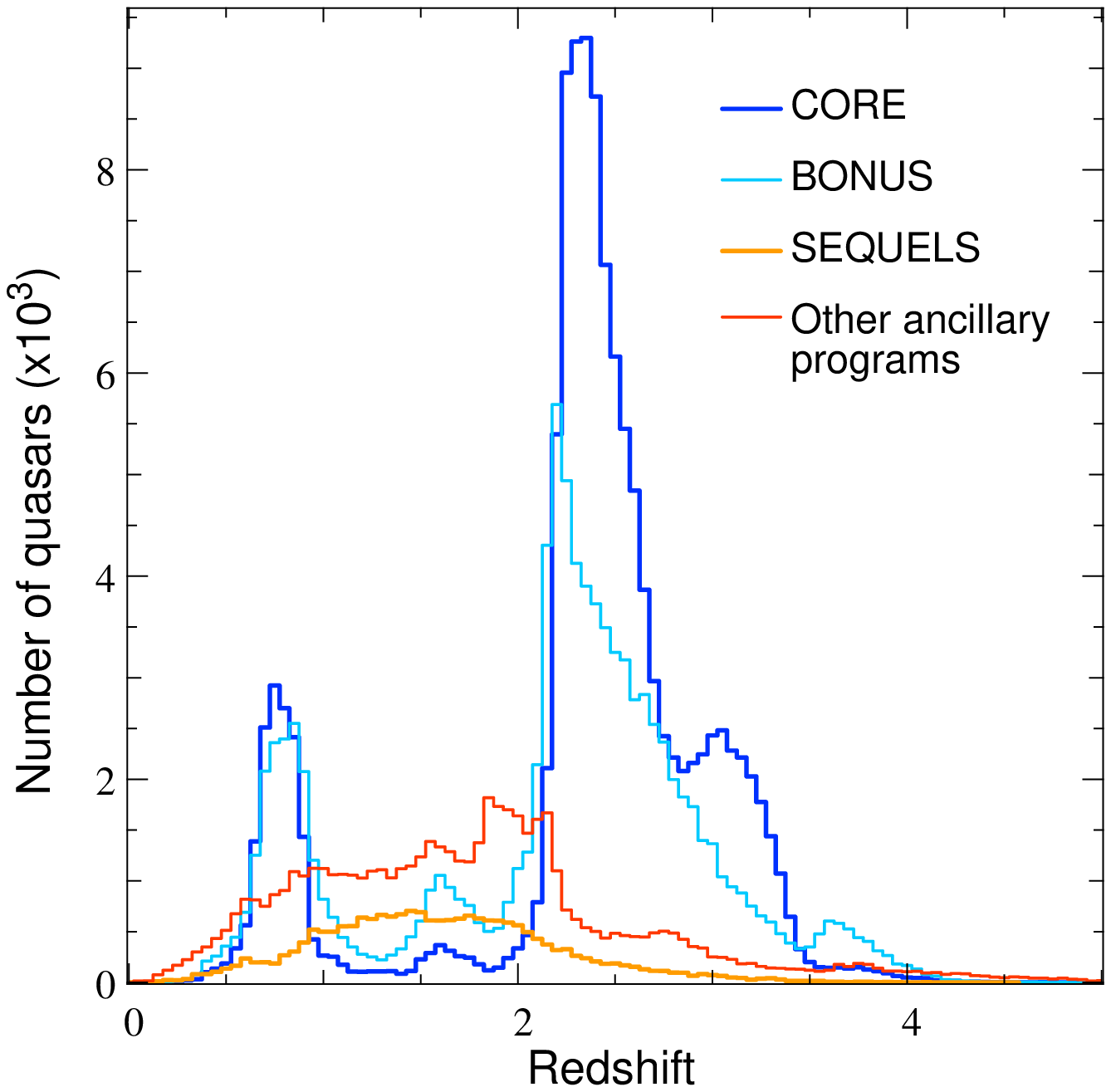}}
		\end{minipage}
		}
	\caption{
	\textit{Left panel: }	
	Redshift distribution of the SDSS-DR12 (thick black histogram), SDSS-DR10 (orange histogram) and SDSS-DR7 (blue histogram) quasars over the redshift range 0-6.
	\textit{Right panel: }
  Redshift distribution of the SDSS-DR12 quasars for the CORE sample (dark blue histogram), BONUS sample (light blue histogram), ancillary programs (red histogram) and SEQUELS (orange histogram).
The bin size of all the histograms is $\Delta z = 0.05$. The two peaks at $z \sim 0.8$ and $z \sim 1.6$ seen in the redshift distribution of the SDSS-DR10 and SDSS-DR12 quasar
samples (left panel) and in the distribution of the CORE and BONUS samples (right panel) are due to known degeneracies in the quasar target
selection \citep[see ][ for details]{ross2012}.
}
	\label{fig:zdistriDR12}
\end{figure*}

SDSS-DR12 quasars have redshifts between $z=0.041$ and $z=6.440$.
The highest redshift quasar in our sample was discovered by \cite{fan2001}.\footnote{\cite{fan2001} reported a slightly different redshift for this quasar (z~=~6.419) based on molecular ISM lines. The DR12Q redshift is solely based on the SDSS-III/BOSS spectrum.}
The redshift distribution of the entire sample is shown in \Fig{zdistriDR12} (left panel; black histogram) and
compared to the SDSS-DR7 \citep[][ blue histogram]{schneider2010} and the SDSS-DR10 \citep[][ orange histogram]{paris2014} quasar catalogs. 
The two peaks at $z \sim 0.8$ and $z \sim 1.6$ seen in the SDSS-DR10 and SDSS-DR12 redshift distributions are due to known degeneracies in the SDSS
color space.
The right panel of \Fig{zdistriDR12} presents the redshift distribution for each main subsample: the CORE sample (dark blue histogram) designed for quasar
demographics and physics studies, the BONUS sample (light blue histogram) whose goal is to maximize the number of $2.15 \leq z \leq 3.5$ quasars, quasars observed
as part of ancillary programs (red histogram) and the SDSS-IV pilot survey, i.e. the SEQUELS sample (orange histogram).

\begin{figure*}[htbp]
	\centering{
		\begin{minipage}{.45\linewidth}
			\centering{\includegraphics[width=75mm]{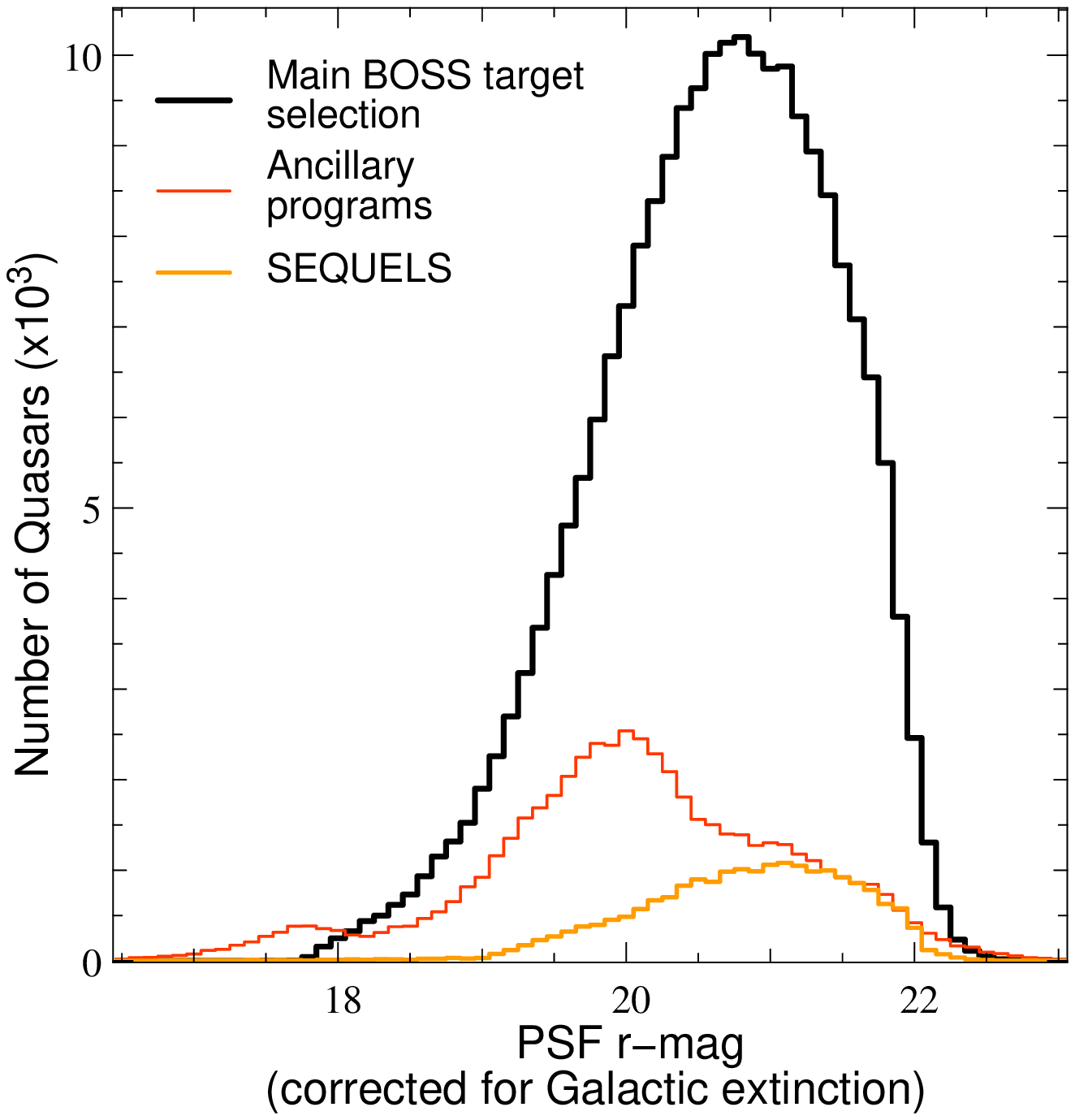}}
		\end{minipage}
		\hfill
		\begin{minipage}{.45\linewidth}
			\centering{\includegraphics[width=75mm]{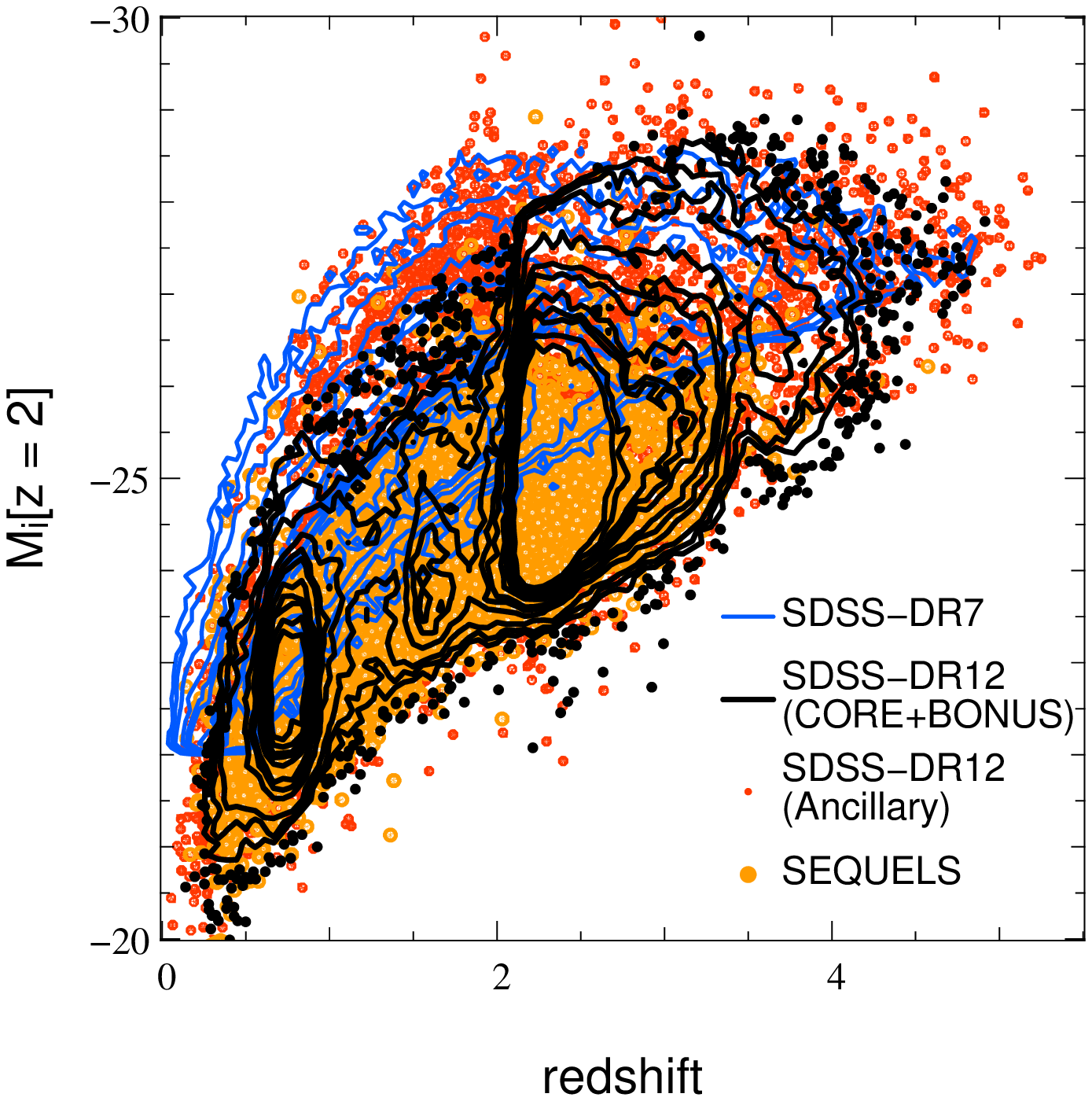}}
		\end{minipage}
		}
	\caption{
	\textit{Left panel: }	
	Distribution of r PSF magnitude (corrected for Galactic extinction) for the BOSS main sample \citep[selected by the algorithms described in ][ thick black histogram]{ross2012},
 ancillary programs (red histogram) and SEQUELS (orange histogram). The bin size is $\Delta r = 0.02$.
	\textit{Right panel: }
  absolute magnitude-redshift plane for SDSS-DR12 CORE+BONUS quasars (black contours), quasars observed as part of SDSS-DR12 ancillary programs (red points), SEQUELS quasars (orange points) and SDSS-DR7 quasars 
 (blue contours).
 The absolute magnitudes assume ${\rm H_0 ~=~ 70 \ km \ s^{-1} \ Mpc^{-1}}$ and the K-correction is given by \cite{richards2006}, who define $K(z = 2) = 0$.
 Contours are drawn at constant point density.
}
	\label{fig:rmag}
\end{figure*}

\begin{figure*}[htbp]
 \centering{\includegraphics[angle=-90,width=150mm]{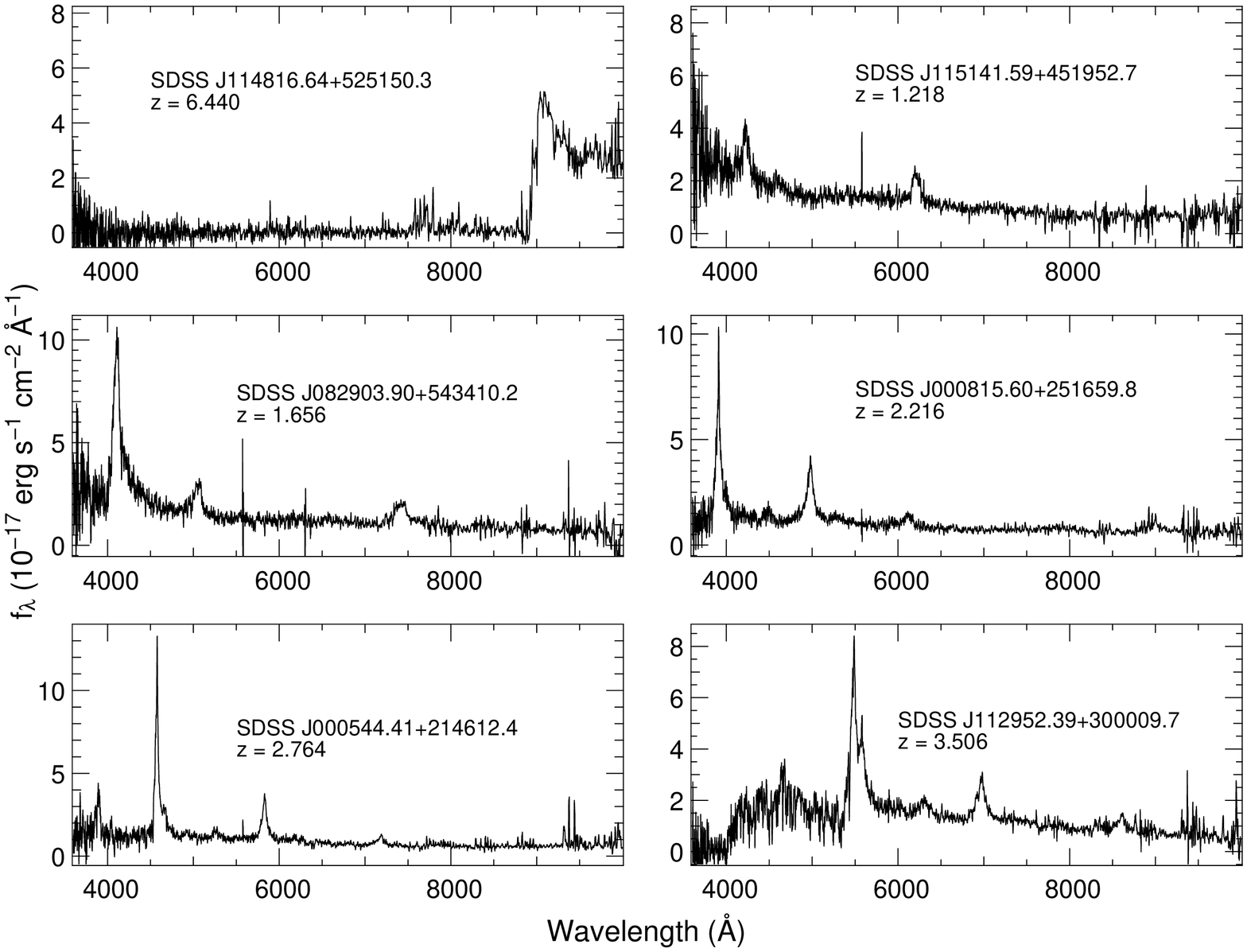}}
 \caption{
  The spectrum of the highest redshift ($z~=~6.440$) quasar observed by BOSS is displayed in the top left panel.
  This quasar was discovered by \cite{fan2001}.
  Typical spectra of SEQUELS quasars are presented at five different redshifts: $z \sim 1.2$ (top right panel) and $z \sim 1.6$ (middle left panel). Typical BOSS quasars are shown at 
  three different redshifts: $z \sim 2.2$ (middle right panel), $z \sim 2.7$ (bottom left panel) and $z \sim 3.5$ (bottom right panel).
  All the spectra are smoothed over 5 pixels.
 }
 \label{fig:exBOSS}
\end{figure*}

The main SDSS-III/BOSS quasar survey is designed to target quasars in the redshift range $2.15 \leq z \leq 3.5$ down to $g = 22$ or $r = 21.85$. The
$r$-band magnitude (corrected for Galactic extinction) distribution of those quasars is displayed in \Fig{rmag} (left panel, black histogram).
Each ancillary program has its own target selection \citep[see e.g., ][]{DR12}, leading to an extremely heterogeneous sample whose $r$-band PSF magnitude distribution is
also shown in \Fig{rmag} (left panel, red histogram).
Lastly, the SDSS-IV pilot survey (SEQUELS) focuses on quasars with $0.9 \leq z \leq 3.5$ down to $g=22$ or $r = 22$ \citep{myers2015}. 
The $r$-band PSF magnitude distribution of the resulting sample is displayed in \Fig{rmag} (left panel, orange histogram).
The right panel of \Fig{rmag} presents the luminosity-redshift plane for the three subsamples contributing to the present catalog: the SDSS-III/BOSS main survey (black contours), quasars observed as part of ancillary programs
(red points) and SEQUELS quasars (orange points). The SDSS-DR7 quasars are plotted for reference in the same plane (blue contours).
Typical quasar spectra are shown in \Fig{exBOSS}.

    \subsection{Known quasars from SDSS-I/II re-observed by SDSS-III/BOSS}

As explained in \Sec{QTS}, a significant fraction of known quasars from DR7Q have been re-observed by
the SDSS-III/BOSS survey.
The main quasar target selection \citep{ross2012} systematically re-targeted DR7Q quasars with $z_{\rm DR7} \geq 2.15$ over the entire survey.
Starting from Year 3, DR7Q quasars with $1.8 \leq z_{\rm DR7} < 2.15$ were also re-observed with the BOSS spectrograph.
Other DR7Q quasars, mostly at lower redshifts, were targeted as part of SDSS-III ancillary programs.
SDSS-III/BOSS spectra were obtained for \numprint{25416} objects listed in DR7Q. A total of \numprint{25275} quasars known from DR7Q are included 
in the present catalog, i.e. are classified as {\tt QSO} or {\tt QSO\_BAL}.
A total of \numprint{59} are not part of the DR12Q catalog due to a lower confidence level of their redshift estimate, i.e., they were classified
as {\tt QSO\_Z?} or {\tt QSO\_BAL\_Z?}, and their current identification 
is included in the DR12 superset file. 
A total of \numprint{14} quasars dropped out of the DR12Q catalog due to bad photometric information and were not visually inspected, and
\numprint{67} quasars have bad spectra, mostly ``empty spectra'', in SDSS-III/BOSS. Our visual inspection is based on SDSS-III/BOSS spectroscopy only and
thus, those spectra were re-classified as {\tt Bad}.
Finally, one object, SDSS J015957.64+003310.5, had its identification changed and was reported to be a ``changing-look quasar'' \citep{lamassa2015,merloni2015}.\\

The DR7Q catalog contains \numprint{19132} quasars with $z \geq 2.15$. In principle, all of these objects should have been re-observed
by SDSS-III/BOSS. However, \numprint{2265} DR7Q quasars lie outside of the SDSS-III/BOSS footprint and hence were not re-targeted.
The other \numprint{16867} quasars have a SDSS-III/BOSS spectrum.
Since the systematic re-observation of DR7Q quasars with $1.8 \leq z < 2.15$ started in Year 3, the completeness for those objects is much lower, with
a total of \numprint{5676} having a SDSS-III/BOSS spectrum (out of \numprint{15432} DR7Q quasars in the same redshift range).

    \subsection{Differences between the SDSS-DR10 and SDSS-DR12 quasar catalogs}
    
The previous SDSS quasar catalog \citep[DR10Q; ][]{paris2014} contained \numprint{166583} quasars, 
\numprint{166520} are included in the DR12Q catalog. Among the \numprint{63} ``missing'' quasars,
\numprint{60} were identified in the galaxy targets and are now part of the supplementary list (see \Sec{suplist}).
As explained in \Sec{defDR12}, we define our parent sample based on quasar candidates from either the main SDSS-III/BOSS survey or
ancillary programs, and we also search for serendipitous quasars among galaxy targets based on the SDSS pipeline identification.
The latter definition depends on the SDSS pipeline version, leading to slight variations between parent samples drawn for each data release. 
However, we keep track of firm identifications from previous pipeline versions and provide them in the
supplementary list.

The identification of two objects reported as quasars in DR10Q changed: one is now classified as {\tt QSO\_?} and the second one as {\tt Bad}.
Those two objects are part of the DR12Q superset with their new identification.
Finally, one actual quasar (SDSS J013918.06+224128.7) dropped out of the main catalog but is part of the supplementary list.
This quasar was re-observed as an emission line galaxy (ELG) candidate. The target selection bits for the two spectra are inconsistent. The best
spectrum as identified by the SDSS pipeline, i.e. having {\tt SPECPRIMARY}~=~1, is associated with the ELG target, but not with the quasar target. It is therefore not selected to be part of our parent sample.

    \subsection{Uniform sample}
    \label{s:uniform}

As in DR9Q and DR10Q, we provide a {\tt uniform} flag (row \#31 of \Tab{DR12Qformat}) in order to identify a homogeneously selected sample of quasars.
The CORE sample has been designed to have a well understood and reproducible selection function. However, its definition varied over the
first year of the survey.
Regions of the sky within which all the algorithms used in the quasar target selection are uniform are called ``Chunks''; their definition is given
in \cite{ross2012}.
After Chunk 12, CORE targets were selected with the XDQSO technique \textit{only} \citep{bovy2011}. 
These objects have {\tt uniform}~=~1.
Quasars in the DR12Q catalog with {\tt uniform}~=~2 are objects that would have been selected by XDQSO if it had been the CORE algorithm prior to Chunk 12.
Objects with {\tt uniform}~=~2 are reasonably complete to what XDQSO would have selected, even prior to Chunk 12. 
 DR12Q only contains information about spectroscopically confirmed quasars and not about all {\em targets}. Hence, care must be taken to draw
a statistical sample from the uniform flag.
See, e.g, the discussion regarding the creation of a statistical sample for clustering measurements in \cite{white2012} and \cite{eftekharzadeh2015}.
Quasars with {\tt uniform}~=~0 are not homogeneously selected CORE targets.

	\subsection{Accuracy of redshift estimates}
	\label{s:redshift}
	
The present catalog contains six different redshift estimates: the visual inspection redshift ({\tt Z\_VI}), the SDSS pipeline redshift ({\tt Z\_PIPE}), a refined PCA redshift ({\tt Z\_PCA}) and three emission-line based redshifts ({\tt Z\_CIV}, {\tt Z\_CIII} and {\tt Z\_MGII}).\\

As described in \Sec{vi_process}, quasar redshifts provided by the SDSS pipeline \citep{bolton2012} are the result of a linear combination of four eigenspectra. The latter were derived from a principal component analysis (PCA) of a sample of quasars with robust redshifts. Hence, this does not guarantee this value to be the most accurate \citep[e.g. ][]{hewett2010,shen2016}.
The SDSS pipeline has a very high success rate in terms of identification and redshift estimate, and in order to correct for the remaining misidentifications and catastrophic redshift failures, we perform the systematic visual inspection of all quasar targets. In case we modified the SDSS pipeline redshift, the visual inspection redshift provides a robust value, set to be at the location of the maximum of the \ion{Mg}{ii} emission line, when this emission line is available in the spectrum. The estimated accuracy is $\Delta z = 0.003$. This value corresponds to the limit below which we are no longer able to see any difference \textit{by eye}.

We also provide a homogeneous automated redshift estimate ({\tt Z\_PCA}) that is based on the result of a PCA performed on a sample of quasars with redshifts measured at the location of the maximum of the \ion{Mg}{ii} emission line \citep[see ][ for details]{paris2012}. The \ion{Mg}{ii} line is chosen as the reference because this is the broad emission line the least affected by systematic shifts with respect to the systemic redshift \citep[e.g. ][]{hewett2010,shen2016}.
In addition to this PCA redshift, we also provide the redshift corresponding to the position of the maximum of the three broad emission lines we fit as part of this catalog: \ion{C}{iv}$\lambda 1550$, \ion{C}{iii]}$\lambda 1909$ and \ion{Mg}{ii}$\lambda 2800$.
\\

\cite{hewett2010} carefully studied the systematic emission-lines shifts in the DR6 quasar sample and computed accurate quasar redshifts by correcting for these shifts.
In DR12Q, there are \numprint{5328} non-BAL quasars that have redshifts measured by \cite{hewett2010} available and for which we also have \ion{Mg}{ii} redshifts that are expected to be the closest estimates to the quasar systemic redshift. \ion{Mg}{ii}-based redshifts are redshifted by 3.3 $\kms$ compared to the \cite{hewett2010} measurements, with a dispersion of 326 $\kms$. This confirms that \ion{Mg}{ii} redshifts provide a reliable and accurate measurement.

\begin{figure}[htbp]
 \centering{\includegraphics[width=75mm]{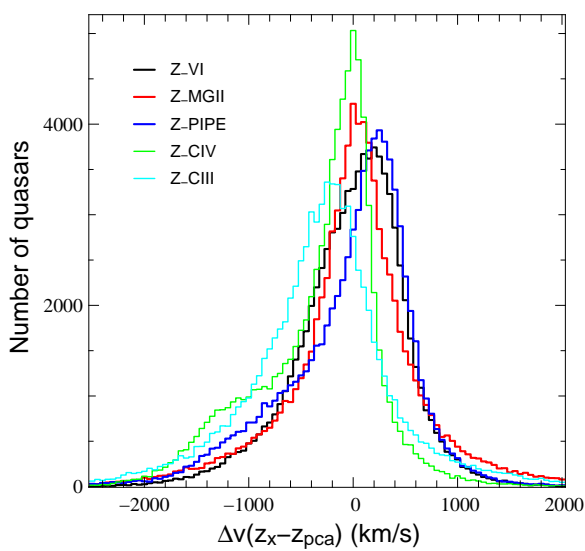}}
 \caption{
	Distribution of velocity differences between the visual inspection redshift ({\tt Z\_VI}, black histogram), \ion{Mg}{ii}-based redshift ({\tt Z\_MGII}, red histogram), SDSS pipeline redshift \citep[][ blue histogram]{bolton2012}, \ion{C}{iv}-based redshift ({\tt Z\_CIV}, green histogram), \ion{C}{iii]}-based redshift ({\tt Z\_CIII}, cyan histogram) and the PCA redshift ({\tt Z\_PCA}).
	The bin size of all the histograms is $\Delta v = 50 \ \kms$.
 }
\label{fig:zestimate}
\end{figure}

\numprint{80791} have the six redshift measurements provided in our catalog available. We compute the redshift difference, expressed as velocities, for {\tt Z\_VI}, {\tt Z\_MGII}, {\tt Z\_PIPE}, {\tt Z\_CIV} and {\tt Z\_CIII} with respect to {\tt Z\_PCA}. We take the latter as our reference because this measurement is uniformally run over the entire DR12Q sample. 
The distributions of redshift differences are shown in \Fig{zestimate}.
\ion{Mg}{ii}-based redshifts are redshifted by 2.1 $\kms$ compared to the PCA redshifts with a dispersion of 680 $\kms$. In average, visual inspection redshifts are blueshifted by 15.3 $\kms$ with respect to the PCA redshifts with a dispersion of 529 $\kms$. The SDSS pipeline estimates are redshifted by 42.7 $\kms$ with a dispersion of 637 $\kms$. Finally, as expected, \ion{C}{iv} and \ion{C}{iii]} redshifts are the most shifted with respect to the PCA redshifts with an average blueshift of 321 $\kms$ and 312 $\kms$ respectively.

    \subsection{Broad Absorption Line quasars}
    \label{s:bal}

As described in \cite{paris2012}, quasars displaying broad absorption lines (BAL) are flagged in the course of the visual inspection of all the quasar targets.
This flag ({\tt BAL\_FLAG\_VI}, Col. \#56 of \Tab{DR12Qformat}) provides qualitative information about the existence of such features \textit{only}. We refer the reader
to Sec. 5 of \cite{paris2012} for a full discussion of the robustness of the visual inspection detection of BALs.
A total of \numprint{29580} quasars have been flagged in the course of visual inspection.

\begin{figure*}[htbp]
	\centering{
		\begin{minipage}{.45\linewidth}
			\centering{\includegraphics[width=75mm]{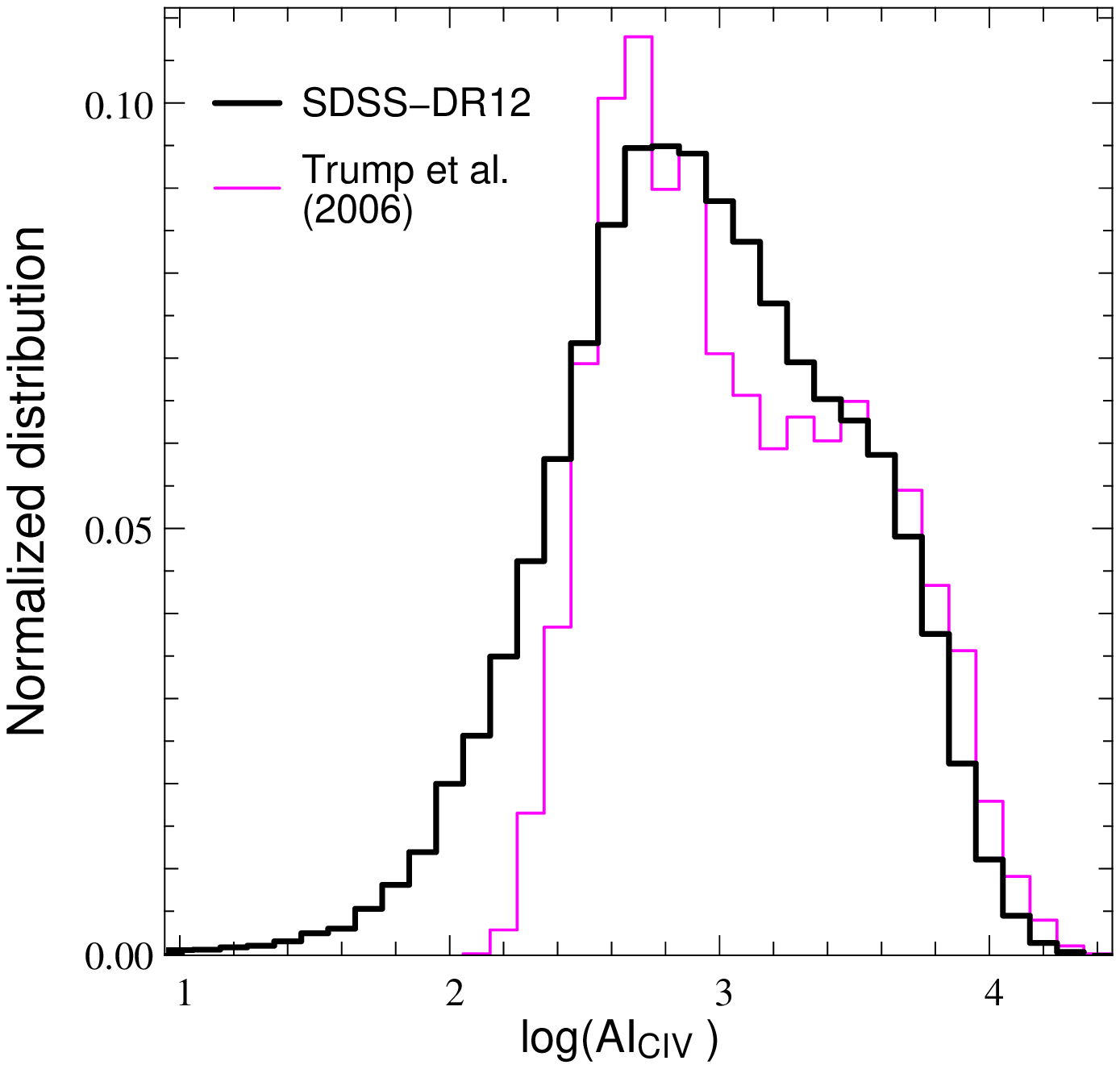}}
		\end{minipage}
		\hfill
		\begin{minipage}{.45\linewidth}
			\centering{\includegraphics[width=75mm]{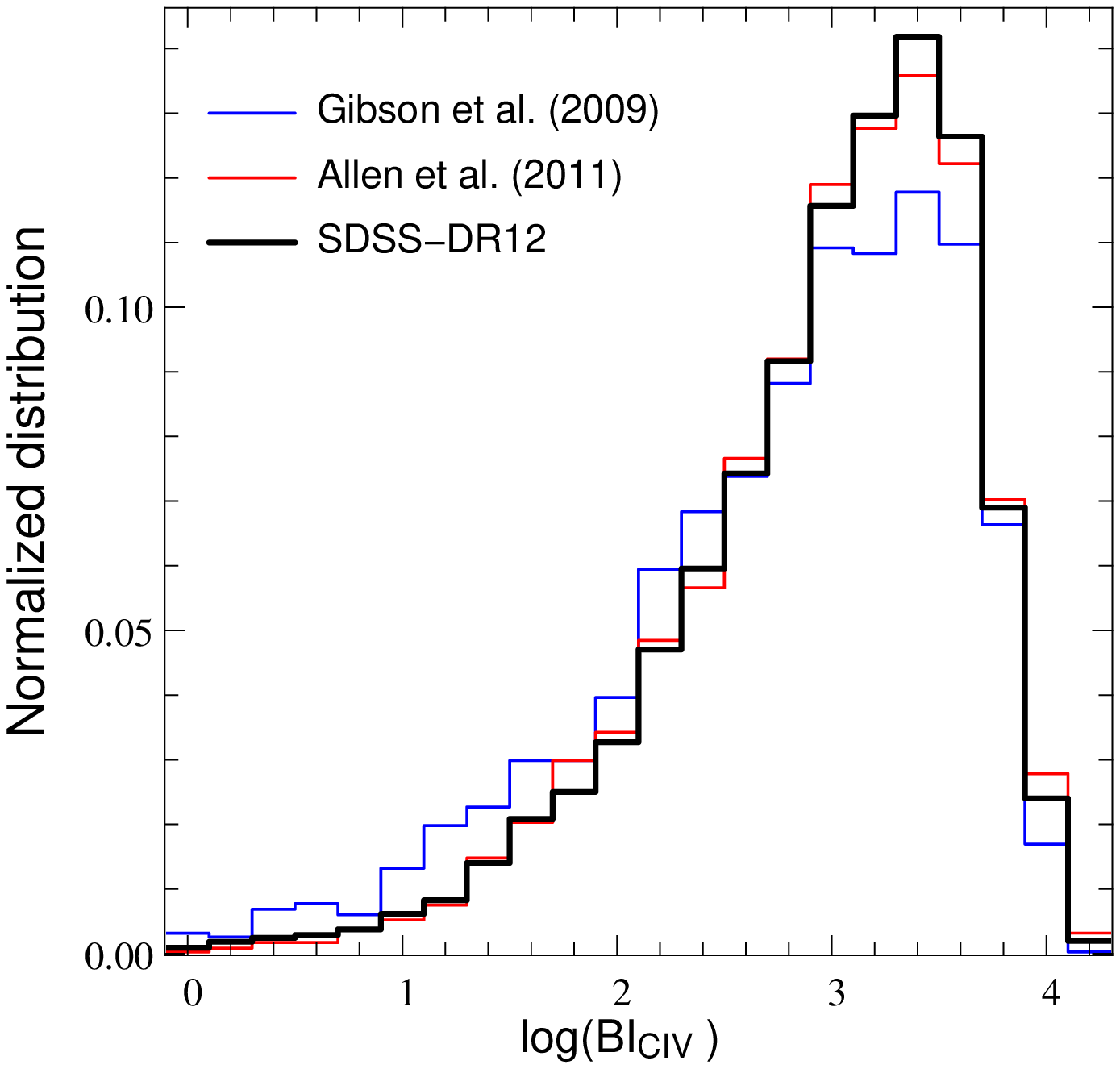}}
		\end{minipage}
		}
	\caption{
	\textit{Left panel: }	
	Distribution of AI for the SDSS-DR3 \citep[magenta histogram; ][]{trump2006} and the SDSS-DR12 (black histogram) quasars.
 Both distributions are normalized to have a surface area equal to 1 for $\log {\rm AI} > 2.5$. 
	\textit{Right panel: }
	Distribution of BI from the DR12Q catalog (black histogram), from the SDSS-DR6 \citep[red histogram; ][]{allen2011} and from the 
SDSS-DR5 \citep[blue histogram; ][]{gibson09}.
	All the distributions are normalized to have their sum equal to 1.
	}
	\label{fig:bal}
\end{figure*}

In order to have a more systematic detection of BALs and 
quantitative measurement of their properties, we automatically search for BAL features and report metrics of common use in the community: 
the BALnicity index \citep[BI; ][]{weymann1991} 
and the absorption index \citep[AI; ][]{hall2002} of the \ion{C}{iv} absorption troughs.
We restrict the automatic search to quasars with {\tt Z\_VI}~$\geq 1.57$ in order to have the full spectral 
coverage of \ion{C}{iv} absorption troughs.
\\ 
The BALnicity index (Col. \#57) is computed bluewards of the \ion{C}{iv} emission line and is defined as:
\begin{equation}
 {\rm BI} ~=~ - \int ^{3000} _{25000} \left[ 1 - \frac{f \left( v \right)}{0.9} \right] C \left( v \right) {\rm d}v ,
 \label{eq:BI_def}
\end{equation}
where $f \left( v \right)$ is the normalized flux density as a function of velocity displacement from the emission-line center. 
The quasar continuum is estimated using a linear combination of four principal components as described in \cite{paris2012}.
$C \left( v \right)$ is initially set to 0 and can take
only two values, 1 or 0. It is set to 1 whenever the quantity $ 1 - f \left( v \right)/0.9$ is continuously positive over an interval of at least \numprint{2000}
$\kms$. It is reset to 0 whenever this quantity becomes negative.
\ion{C}{iv} absorption troughs wider than \numprint{2000} $\kms$ are detected in the spectra of \numprint{16442} quasars. 
The number of lines of sight in which {\tt BI\_CIV} is positive is lower than the number of visually flagged BAL for several reasons:
(i) the visual inspection provides a qualitative flag for absorption troughs that do not necessarily meet the requirements defined in \Eq{BI_def},
(ii) the visual inspection is not restricted to any search window, and 
(iii) the visual inspection flags rely on absorption troughs due to other species in addition to \ion{C}{iv} (especially \ion{Mg}{ii}).
The distribution of BI for \ion{C}{iv} troughs from DR12Q is plotted in the right panel of \Fig{bal} (black histogram) and is compared 
to previous works by \citeauthor{gibson09} \citeyearpar[blue histogram]{gibson09} performed on DR5Q \citep{schneider2007} and 
by \citeauthor{allen2011} \citeyearpar[red histogram]{allen2011} who searched for BAL quasars in quasar spectra released as part
of SDSS-DR6 \citep{DR6}.
The three distributions are normalized. 

The overall shapes of the three distributions are similar. 
The BI distribution from \cite{gibson09} exhibits an excess of low-BI values compared to \cite{allen2011} 
and this work.  The most likely explanation is the difference in the quasar emission modeling. 
\cite{allen2011} used a non-negative matrix factorization (NMF) to estimate the unabsorbed
flux, which produces a quasar emission line shape akin to the one we obtain with PCA. \cite{gibson09} modeled the 
quasar continuum with a reddened power-law and strong emission lines with Voigt profiles. 
Power-law like continua tend to underestimate the actual quasar emission and hence, the resulting BI values tend to be lower
than the one computed when the quasar emission is modeled with NMF or PCA methods.
\\

For weaker absorption troughs, we compute the absorption index (Col. \#59) as defined by \cite{hall2002}:
\begin{equation}
 {\rm AI} ~=~ - \int ^{0} _{25000} \left[ 1 - \frac{f \left( v \right)}{0.9} \right] C \left( v \right) {\rm d}v ,
 \label{eq:AI_def}
\end{equation}
where $f \left( v \right)$ is the normalized continuum and $C \left( v \right)$ has the same definition as in \Eq{BI_def} except that the threshold to set $C$ to 1
is reduced to \numprint{450} $\kms$. Following \cite{trump2006}, we calculate the reduced $\chi ^2$ for each trough, i.e. where $C=1$:
\begin{equation}
 \chi ^2 _{\rm trough} ~=~ \sum \frac{1}{{\rm N}}\left( \frac{1 - f \left( v \right)}{\sigma} \right) ^2,
 \label{eq:chi2_trough}
\end{equation}
where ${\rm N}$ is the number of pixels in the trough, $f \left( v \right)$ is the normalized flux and $\sigma$ the estimated rms noise for each pixel.
The greater the value of $\chi ^2 _{\rm trough}$, the more likely the trough is not due to noise.
Following the recommendations of \cite{trump2006}, we report AIs when at least one \ion{C}{iv} absorption trough is detected 
with $\chi ^2 _{\rm trough} \geq 10$.
A total of \numprint{48863} quasars satisfy this definition.
The distribution of AI with $\chi ^2 _{\rm trough} \geq 10$ is shown in the left panel of \Fig{bal} (black histogram)
and is compared to the results of \citeauthor{trump2006} \citeyearpar[magenta histogram]{trump2006}. Both distributions are normalized to have a surface area equal to 1 for $\log {\rm AI} > 2.5$.
The distributions are similar. The main difference between the two distributions is the existence of a tail at low AI in DR12Q. This is due to the
modified definition for AI used in \cite{trump2006}: the 0.9 factor in \Eq{AI_def} was removed and the detection threshold
was larger (\numprint{1000} $\kms$ instead of \numprint{450} $\kms$ here).

We also report the number of troughs detected with widths larger than \numprint{2000} $\kms$ (Col. \#62 in \Tab{DR12Qformat})
and larger than \numprint{450} $\kms$ (Col. \#65) together with the velocity ranges
over which these troughs are detected (Cols.  \#63-64 and  \#66-67, respectively).

BAL information reported in the DR12Q catalog is computed over all the absorption troughs that are detected.
However, out of \numprint{29580} quasars with {\tt BAL\_FLAG\_VI}~=~1, i.e. flagged during the visual inspection process, \numprint{16693}
have more than one absorption trough larger than \numprint{450} $\kms$ detected in the spectrum (out of \numprint{21444} with
AI$>$0),
and \numprint{2564} with more than one absorption trough larger than \numprint{2000} $\kms$ (out of \numprint{15044} with BI$>$0).
We provide detailed ``trough-by-trough''
information for all {\tt BAL\_FLAG\_VI}~=~1 quasars in a separate file whose format is described in \Tab{bal_format}.
For each detected absorption trough, we measure the velocity range 
in which the normalized flux density\footnote{The normalized flux density is defined as $F_{\rm obs}/F_{\rm QSO}$ where $F_{\rm obs}$ is the observed flux density and $F_{\rm QSO}$ is the modeled quasar
emission.} is measured to be lower than 0.9, given the position of its minimum and the value of the
normalized flux density at this location.

 \begin{table*}
  \centering                        
  \begin{tabular}{c l c l}        
  \hline\hline
  Column & Name & Format & Description\\
  \hline\hline
  1    & SDSS\_NAME                      &  STRING      & SDSS-DR12 designation  hhmmss.ss+ddmmss.s (J2000)\\
  2    & RA                              &  DOUBLE      & Right Ascension in decimal degrees (J2000)\\	
  3    & DEC                             & DOUBLE       & Declination in decimal degrees (J2000)\\
  4    & THING\_ID                       &  INT32       & Thing\_ID (unique SDSS source identifier)\\
  5    & PLATE                           & INT32        & Spectroscopic Plate number \\
  6    & MJD                             & INT32        & Spectroscopic MJD \\
  7    & FIBERID                         & INT32        & Spectroscopic Fiber number \\
  \hline
  8    & Z\_VI                           &  DOUBLE      & Redshift from visual inspection \\
  9    & Z\_PIPE                         &  DOUBLE      & Redshift from BOSS pipeline \\
  10   & ERR\_ZPIPE                      &  DOUBLE      & Error on BOSS pipeline redshift \\
  11   & ZWARNING                        & INT32        & ZWARNING flag  \\
  12   & Z\_PCA                          & DOUBLE       & PCA redshift \\
  13   & ERR\_ZPCA                       & DOUBLE       & Error on PCA redshift \\
  \hline
  14   & SDSS\_MORPHO                    &   INT32      & SDSS morphology flag 0 = point source 1 = extended \\
  15   & BOSS\_TARGET1                   &  INT64       & BOSS target flag for main survey  \\	
  16   & ANCILLARY\_TARGET1              &   INT64      & BOSS target flag for ancillary programs \\
  17   & ANCILLARY\_TARGET2              &   INT64      & BOSS target flag for ancillary programs  \\
  18   & EBOSS\_TARGET0                  &   INT64      & Target selection flag for eBOSS pilot survey \\
  \hline
  19   & PSFFLUX                         & FLOAT[5]     & flux in the $u$,$g$,$r$,$i$ and $z$-bands (not corrected for Galactic extinction)\\
  20   & IVAR\_PSFLUX                    & FLOAT[5]     & inverse variance of $u$,$g$,$r$,$i$ and $z$ fluxes \\
  21   & PSFMAG                          & FLOAT[5]     & PSF magnitudes in $u$,$g$,$r$,$i$ and $z$-bands (not corrected for Galactic \\
       &                                 &              & extinction)\\
  22   & ERR\_PSFMAG                     & FLOAT[5]     & error in $u$,$g$,$r$,$i$ and $z$ PSF magnitudes\\
  23   & GAL\_EXT                        & FLOAT[5]     & Galactic extinction in the 5 SDSS bands \citep[from ][]{schlegel1998} \\
  \hline
  24   & SNR\_SPEC                       &    FLOAT      & Median signal-to-noise ratio per pixel over the whole spectrum\\
  25   & SNR\_1700                       &    FLOAT      & Median signal-to-noise ratio per pixel in the window \numprint{1650} - \numprint{1750}\AA\ \\
       &                                 &               & (rest frame)\\
  26   & SNR\_3000                       &    FLOAT      & Median signal-to-noise ratio per pixel in the window \numprint{2950} - \numprint{3050}\AA\ \\
       &                                 &               & (rest frame)\\
 
  \hline
  27   & BI\_CIV                         & DOUBLE       & Balnicity Index (BI) \\
  28   & ERR\_BI\_CIV                    & DOUBLE       & Error on Balnicity index \\
  29   & NCIV\_2000                      & INT32        & Number of troughs wider than \numprint{2000} ${\rm km \ s^{-1}}$ \\
  30   & VMIN\_CIV\_2000                 & DOUBLE[5]    & Minimum velocity of each detected absorption trough (Col. \#29)\\
  31   & VMAX\_CIV\_2000                 & DOUBLE[5]    & Maximum velocity of each detected absorption trough (Col. \#29)\\
  32   & POSMIN\_CIV\_2000               & DOUBLE[5]    & Position of the minimum of each absorption trough \\
  33   & FMIN\_CIV\_2000                 & DOUBLE[5]    & Normalized flux density at the minimum of each absorption trough \\
  34   & AI\_CIV                         & DOUBLE       & Absorption index (AI)\\
  35   & ERR\_AI\_CIV                    & DOUBLE       & Error on absorption index \\
  36   & NCIV\_450                       & INT32        & Number of absorption trough wider than \numprint{450} ${\rm km \ s^{-1}}$ \\
  37   & VMIN\_CIV\_450                  & DOUBLE[17]    & Minimum velocity of each detected absorption trough (Col. \#36) \\
  38   & VMAX\_CIV\_450                  & DOUBLE[17]    & Maximum velocity of each detected absorption trough (Col. \#36) \\
  39   & POSMIN\_CIV\_450                & DOUBLE[17]    & Position of the minimum of each absorption trough \\
  40   & FMIN\_CIV\_450                  & DOUBLE[17]    & Normalized flux density at the minimum of each absorption trough \\
  \hline
  \hline
  \end{tabular}
  \caption{Format of the binary FITS file that contains ``trough-by-trough'' information for BAL quasars.}
  \label{t:bal_format}
  \end{table*}

    \subsection{Multi-wavelength cross-correlation}

We provide multi-wavelength matching of DR12Q quasars to several surveys:    
the FIRST radio survey \citep{becker1995}, the Galaxy Evolution Explorer 
\citep[GALEX,][]{martin2005} survey in the UV, the Two Micron All Sky Survey \citep[2MASS,][]{cutri2003,skrutskie2006},  the UKIRT Infrared Deep 
Sky Survey \citep[UKIDSS;][]{lawrence2007}, 
the Wide-Field Infrared Survey \citep[WISE,][]{wright2010}, the ROSAT All-Sky Survey \citep[RASS;][]{voges1999,voges2000}, and the 
fourth data release of the Third XMM-Newton Serendipitous
Source Catalog \citep{watson2009}.

	\subsubsection{FIRST}
	\label{s:FIRST}

As for the previous SDSS-III/BOSS quasar catalogs, we matched the DR12Q quasars to the latest FIRST catalog \citep[March 2014; ][]{becker1995} using a 2\arcsec \ matching radius.
We report the flux peak density at 20 cm and the signal-to-noise ratio of the detection.
Among the DR12Q quasars, \numprint{29671} lie outside of the FIRST footprint and have their {\tt FIRST\_MATCHED} flag set to -1.

The SDSS-III/BOSS quasar target selection \citep{ross2012} automatically includes matches to the FIRST sources from a {\em previous} (July 2008) 
version of the FIRST catalog  that have $(u-g) > 0.4$. 
This additional color cut is set to avoid contamination by low-redshift quasars.

A total of \numprint{10221} quasars have FIRST counterparts in DR12Q.
We estimate the fraction of chance superposition by offsetting the declination of DR12Q quasars by 200\arcsec . We then re-match to the FIRST source catalog. We conclude that there are about 0.2\% of false positives in the DR12Q-FIRST matching.

	\subsubsection{The Galaxy Evolution Explorer (GALEX)}
	\label{s:GALEX}

As for DR10Q, GALEX \citep{martin2005} images are force-photometered (from GALEX Data Release 5) at the SDSS-DR8 centroids \citep{DR8}, such that low S/N point-spread function fluxes of objects not detected by GALEX are recovered,  
for both the FUV (\numprint{1350}-\numprint{1750} \AA ) and NUV (\numprint{1750}-\numprint{2750} \AA ) bands when available.
A total of \numprint{197781} quasars are detected in the NUV band, \numprint{158474} in the FUV band and \numprint{129090} have non-zero fluxes in both bands.

	\subsubsection{The Two Micron All Sky Survey (2MASS)}
	\label{s:2mass}

We cross-correlate DR12Q with the All-Sky Data Release Point Source catalog \citep{skrutskie2006} using a matching radius of 2\arcsec.
We report the Vega-based magnitudes in the J, H and K-bands and their error together with the signal-to-noise ratio of the detections.
We also provide the value of the 2MASS flag rd\_flg[1], which defines the peculiar values of the magnitude and its error for each band\footnote{see http://www.ipac.caltech.edu/2mass/releases/allsky/doc/explsup.html for more details}.

There are \numprint{471} matches in the catalog. This number is quite small compared with the number of DR12Q quasars because the sensitivity of 2MASS is much less than that of SDSS.
Applying the same method as described in \Sec{FIRST}, we estimate that 0.8\% of the matches are false positives.

	\subsubsection{The Wide-Field Infrared Survey (WISE)}
	\label{s:wise}

Previous SDSS-III/BOSS quasar catalogs \citep{paris2012,paris2014} were matched to the WISE All-Sky Source catalog\footnote{http://wise2.ipac.caltech.edu/docs/release/allsky/}
\citep{wright2010}.
For this version of the quasar catalog, we matched the DR12Q to the newly released AllWISE Source Catalog\footnote{http://wise2.ipac.caltech.edu/docs/release/allwise/} \citep{wright2010,mainzer2011}
that has enhanced photometric sensitivity and accuracy, and improved astrometric precision.
Our procedure is the same as in DR9Q and DR10Q, with a matching radius of 2.0\arcsec\ which provides a low level of false positive matches \citep[see e.g. Fig. 4 of][]{krawczyk2013}.
There are \numprint{190408} matches from the AllWISE Source Catalog.
Following the procedure described in \Sec{FIRST}, we estimate the rate of false positive matches to be about 2\%.

We report the magnitudes, their associated errors, the S/N of the detection and reduced $\chi ^2$ of the profile-fitting in 
the four WISE bands centered at wavelengths of 3.4, 4.6, 12 and 22~$\mu$m. 
These magnitudes are in the Vega system, and are measured with profile-fitting photometry.
We also report the WISE catalog contamination and confusion flag, {\tt cc\_flags}, and their photometric quality flag, {\tt ph\_qual}.
As suggested on the WISE ``Cautionary Notes" page\footnote{http://wise2.ipac.caltech.edu/docs/release/allsky/expsup/sec1\_4b.html \#unreliab}, we
recommend using only those matches with {\tt cc\_flags} = ``0000'' to exclude objects that are flagged as spurious detections of image artifacts in any band.
Full details about quantities provided in the AllWISE Source Catalog can be found on their online documentation\footnote{http://wise2.ipac.caltech.edu/docs/release/allsky/expsup/ sec2\_2a.html}.

	\subsubsection{UKIDSS}
	\label{s:ukidss}

As for DR10Q, near infrared images from the UKIRT Infrared Deep Sky Survey \citep[UKIDSS; ][]{lawrence2007} are force-photometered
at the SDSS-DR8 centroids \citep{DR8}.

We provide the fluxes and their associated errors, expressed in ${\rm W \ m^{-2} \ Hz^{-1}}$, in the Y, J, H and  K bands.
The conversion to the Vega magnitudes, as used in 2MASS, is given by the formula:
\begin{equation}
	{\rm mag _{X}} = -2.5 \times \log \frac{f_{\rm X}}{f_{\rm 0, X} \times 10^{-26}},
\end{equation}
where X denotes the filter and the zero-point values $f_{\rm 0, X}$ are 2026, 1530, 1019 and 631 for the Y, J, H and K bands respectively.

A total of \numprint{78783} quasars are detected in at least one of the four bands Y, J, H or K.
\numprint{78079} objects are detected in the Y band, \numprint{77761} in the J band, \numprint{77726} in H band, \numprint{78179} in the K band and \numprint{75987} objects have 
non-zero fluxes in the four bands. Objects with zero fluxes lie outside the UKIDSS footprint.
The UKIDSS limiting magnitude is ${\rm K} \sim 18$ (for the Large Area Survey) while the 2MASS limiting magnitude in the same band is $\sim 15.3$.
This difference in depth between the two surveys explains the large difference in the numbers of matches with DR12Q.

	\subsubsection{ROSAT all sky survey}
	\label{s:rosat}

As we did for the previous SDSS-III/BOSS quasar catalogs, we matched the DR12Q quasars to the ROSAT all sky survey Faint \citep{voges2000} and Bright \citep{voges1999} source 
catalogues with a matching radius of 30\arcsec.
Only the most reliable detections are included in our catalog: when the quality detection is flagged as potentially problematic, we do not include 
the match.
A total of \numprint{1463} quasars are detected in one of the RASS catalogs.
As for the cross-correlations described above, we estimate that 2.1\% of the RASS-DR12Q matches are due to chance superposition.
	
	\subsubsection{XMM-Newton}
	\label{s:xmm}

DR12Q was cross-correlated with the fourth data release of the Third XMM-Newton Serendipitous
Source Catalog\footnote{http://xmmssc-www.star.le.ac.uk/Catalogue/xcat\_public\_3XMM-DR4.html} (3XMM-DR4)
using a standard 5.0\arcsec \ matching radius.
The 3XMM-DR4 catalog benefits from an increase of the number of public observations leading to an increase of $\sim$40\% of the number
of unique sources. Furthermore, significant improvements of the XMM-Newton science analysis software and calibration
allow the detection of fainter sources.
Thanks to these changes in the XMM-Newton Serendipitous Source Catalog, the overall number of matches increased by $\sim$50\%.
For each of the \numprint{5354} DR12Q quasars with XMM-Newton counterparts, we report the total flux (0.2--12 keV), and associated error, from the
three XMM-Newton CCDs (MOS1, MOS2, PN).
In the case of multiple XMM-Newton observations, the one with the longest exposure time was used to compute the total flux.
We also report the soft (0.2--2 keV), hard (4.5--12 keV) and total (0.2--12 keV) fluxes, and associated errors, that were
computed as the weighted average of all the detections in the three cameras.
Corresponding observed X-ray luminosities are computed in each band and are not absorption corrected.
All fluxes and errors are expressed in ${\rm erg \ cm^{-2} \ s^{-1}}$ and luminosities are computed using the visual inspection redshift ({\tt Z\_VI}). 
Finally, in the case of no detection or detection with significant errors (less than $1\sigma $ detections), we provide an 
upper limit for the flux in the hard band (2--10 keV).
Such sources have the flag {\tt LX2\_10\_UPPER} set to -1.

    \subsection{Variability}
    \label{s:var}

Photometric variability has proven to be an efficient method to distinguish quasars from stars even in redshift ranges where their colors overlap  \citep[e.g. ][]{palanque2011,palanque2013a,peters2015}.
The variability of astronomical sources can be characterized by their ``structure function''  that is a measurement of the
amplitude of the observed variability as a function of the time delay between two observations \citep[e.g. ][]{schmidt2010}.
This function is modeled as a power-law parametrized in terms of $A$, the mean amplitude of the variation on a one-year timescale (in the observer's reference
frame) and $\gamma$, the logarithmic slope of the variation amplitude with respect to time.
With $\Delta m _{ij}$ defined as the difference between the magnitudes of the source at time $t_i$ and $t_j$, and assuming an underlying Gaussian distribution
of $\Delta m$ values, the model predicts an evolution of the variance $\sigma ^2 \left( \Delta m \right)$ with time according to
\begin{equation}
 \sigma ^2 \left( \Delta m \right) ~=~ \left[ A \left( \Delta t_{ij} \right) ^{\gamma} \right] ^2 + \left( \sigma ^2 _i + \sigma ^2 _j \right),
 \label{eq:str_func}
\end{equation}
where $\sigma _i$ and $\sigma _j$ are the photometric errors at time $t_i$ and $t_j$.
Quasars are expected to lie at high $A$ and $\gamma$, non-variable stars to lie near $A ~=~ \gamma ~=~ 0$ and variable stars to have $\gamma \sim 0$ even if
$A$ can be large.
In addition, variable sources are expected to deviate greatly in a $\chi ^2$ sense from a model with constant flux.\\

DR12Q quasars lying in the Stripe 82 region, i.e. with $300 \degree < \alpha _{\rm J2000} < 360 \degree$ or $0 \degree <  \alpha _{\rm J2000} < 50 \degree$ and $-1.25 \degree < \delta _{\rm J2000} < +1.25 \degree$,
have typically 60 epochs of imaging reported in DR9 \citep{DR9}.
In the rest of the SDSS-III/BOSS footprint, photometric data from SDSS-DR9 \citep{DR9} and the Palomar Transient Factory \citep[PTF; ][]{rau2009,law2009} were combined to construct light curves,
providing between 3 and 10 epochs. A ``PTF epoch'' in this context actually corresponds to coadded imaging data 
from a few tens of single-epoch PTF images, up to a thousand in the case of some frequently observed PTF fields.
The duration of one ``PTF epoch'' ranges from several months to one year.
Combined PTF+SDSS light curves, and the variability related observables, are consistently measured in the PTF R-band filter, which is
related to SDSS photometric system by $R \simeq r + \alpha \left( r - i \right)$ with $\alpha = 0.20-0.22$ \citep{ofek2012}.

As part of DR12Q, we provide the structure function parameters $A$ ({\tt VAR\_A}) and $\gamma$ ({\tt VAR\_GAMMA}) as defined in \Eq{str_func} for each object with variability data.
We also give the reduced $\chi ^2$ ({\tt VAR\_CHI2}) when each light curve is fitted with a constant, i.e., when no variability is assumed.
The value of the {\tt VAR\_MATCHED} flag is the number of photometric epochs used to estimate the quasar variability. 
A total of \numprint{143359} quasars have variability data available.
When no variability information
is available, this flag is set to 0.

%
\section{Description of the DR12Q catalog}
\label{s:description}

The DR12Q catalog is available as a binary FITS table file at the SDSS public website\footnote{www.sdss.org/dr12/algorithms/boss-dr12-quasar-catalog}.
All the required documentation (format, name, unit of each column) is provided in the FITS header, and 
it is also summarized in \Tab{DR12Qformat}.\\

\longtab[1]{
\begin{longtable}{clcl}
\caption{\label{t:DR12Qformat} DR12Q catalog format}\\
\hline\hline
Column & Name & Format & Description$^a$ \\
\hline
\endfirsthead
\caption{Continued.} \\
\hline\hline
Column & Name  &  Format & Description \\
\hline
\endhead
\hline
\endfoot
\hline
\endlastfoot
%
1    & SDSS\_NAME                      &  STRING       & SDSS-DR12 designation  hhmmss.ss+ddmmss.s (J2000)\\
2    & RA                              &  DOUBLE       & Right Ascension in decimal degrees (J2000)\\
3    & DEC                             & DOUBLE        & Declination in decimal degrees (J2000)\\
4    & THING\_ID                       &  INT32        & Thing\_ID (unique identifier of SDSS sources)\\
5    & PLATE                           & INT32         & Spectroscopic Plate number \\
6    & MJD                             & INT32         & Spectroscopic MJD \\
7    & FIBERID                         & INT32         & Spectroscopic Fiber number \\
\hline
%
%
8    & Z\_VI                           &  DOUBLE       & Redshift from visual inspection \\
9    & Z\_PIPE                         &  DOUBLE       & Redshift from BOSS pipeline \\
10   & ERR\_ZPIPE                      &  DOUBLE       & Error on BOSS pipeline redshift \\
11   & ZWARNING                        & INT32         & ZWARNING flag  \\
12   & Z\_PCA                          & DOUBLE        & Refined PCA redshift \\
13   & ERR\_ZPCA                       & DOUBLE        & Error on refined PCA redshift \\
14   & PCA\_QUAL                       & DOUBLE        & Estimator of the PCA continuum quality\\
15   & Z\_CIV                          & DOUBLE        & Redshift of \CIV\ emission line \\
16   & Z\_CIII                         & DOUBLE        & Redshift of \CIII\ emission complex \\
17   & Z\_MGII                         & DOUBLE        & Redshift of \MgII\ emission line\\
\hline
%
%
18   & SDSS\_MORPHO                    &   INT32       & SDSS morphology flag 0 = point source 1 = extended \\
19   & BOSS\_TARGET1                   &  INT64        & BOSS target flag for main survey  \\
20   & ANCILLARY\_TARGET1              &   INT64       & BOSS target flag for ancillary programs \\
21   & ANCILLARY\_TARGET2              &   INT64       & BOSS target flag for ancillary programs  \\
22   & EBOSS\_TARGET0                  &   INT64       & Target selection flag for the SDSS-IV/eBOSS pilot survey \\
23   & NSPEC\_BOSS                     &   INT32       & Number of additional spectra available in SDSS-III/BOSS \\
24   & PLATE\_DUPLICATE                &   INT32[32]   & Spectroscopic plate number for each duplicate spectrum \\
25   & MJD\_DUPLICATE                  &   INT32[32]   & Spectroscopic MJD for each duplicate spectrum \\
26   & FIBERID\_DUPLICATE              &   INT32[32]   & Spectroscopic fiber number for each duplicate spectrum \\
27   & SDSS\_DR7                       & INT32         & 1 if the quasar is known from DR7\\ 
28   & PLATE\_DR7                      & INT32         & SDSS-DR7 spectroscopic Plate number if the quasar is known from DR7 \\
29   & MJD\_DR7                        &  INT32        & SDSS-DR7 spectroscopic MJD  if the quasar is known from DR7\\
30   & FIBERID\_DR7                    & INT32         & SDSS-DR7 spectroscopic Fiber number  if the quasar is known from DR7\\
31   & UNIFORM                         & INT32         & Uniform sample flag \\
32   & ALPHA\_NU                       &  FLOAT        & Spectral index measurement $\alpha _{\nu}$ \\
\hline
%
%
33   & SNR\_SPEC                       &    FLOAT      & Median signal-to-noise ratio over the whole spectrum\\
34   & SNR\_DUPLICATE                  &   FLOAT[32]   & Median signal-to-noise ratio over the whole spectrum of duplicate \\
     &                                 &               & SDSS-III/BOSS spectra when available \\
35   & SNR\_1700                       &    FLOAT      & Median signal-to-noise ratio in the window \numprint{1650} - \numprint{1750}\AA\ (rest frame)\\
36   & SNR\_3000                       &    FLOAT      & Median signal-to-noise ratio in the window \numprint{2950} - \numprint{3050}\AA\ (rest frame)\\
37   & SNR\_5150                       &    FLOAT      & Median signal-to-noise ratio in the window \numprint{5100} - \numprint{5250}\AA\ (rest frame)\\
38   & FWHM\_CIV                       &   DOUBLE      & FWHM of \CIV\ emission line in ${\rm km \ s^{-1}}$ \\
39   & BHWHM\_CIV                      &    DOUBLE     & Blue HWHM of \CIV\ emission line in ${\rm km \ s^{-1}}$ \\
40   & RHWHM\_CIV                      &    DOUBLE     & Red HWHM of \CIV\ emission line in ${\rm km \ s^{-1}}$ \\
41   & AMP\_CIV                        &   DOUBLE      & Amplitude of \CIV\ emission line in units of median rms pixel noise \\
42   & REWE\_CIV                       &   DOUBLE      & Rest frame equivalent width of \CIV\ emission line in \AA \\
43   & ERR\_REWE\_CIV                  &  DOUBLE       & Uncertainty on the rest frame equivalent width of \CIV\ emission line in \AA \\
44   & FWHM\_CIII                      &     DOUBLE    & FWHM of \CIII\ emission complex in ${\rm km \ s^{-1}}$ \\
45   & BHWHM\_CIII                     &     DOUBLE    & Blue HWHM of \CIII\ emission line in ${\rm km \ s^{-1}}$ \\
46   & RHWHM\_CIII                     &     DOUBLE    & Red HWHM of \CIII\ emission line in ${\rm km \ s^{-1}}$ \\
47   & AMP\_CIII                       &    DOUBLE     & Amplitude of \CIII\ emission complex in units of median rms pixel noise\\
48   & REWE\_CIII                      &      DOUBLE   & Rest frame equivalent width of \CIII\ emission line in \AA \\
49   & ERR\_REWE\_CIII                 &  DOUBLE       & Uncertainty on the rest frame equivalent width of \CIII\ emission \\
     &                                 &               & complex in \AA \\
50   & FWHM\_MGII                      &    DOUBLE     & FWHM of \MgII\ emission line in ${\rm km \ s^{-1}}$ \\
51   & BHWHM\_MGII                     &      DOUBLE   & Blue HWHM of \MgII\ emission line in ${\rm km \ s^{-1}}$ \\
52   & RHWHM\_MGII                     &      DOUBLE   & Red HWHM of \MgII\ emission line in ${\rm km \ s^{-1}}$ \\
53   & AMP\_MGII                       &     DOUBLE    & Amplitude of \MgII\ emission line in units of median rms pixel noise \\
54   & REWE\_MGII                      &      DOUBLE   & Rest frame equivalent width of \MgII\ emission line in \AA \\
55   & ERR\_REWE\_MGII                 &  DOUBLE       & Uncertainty on the rest frame equivalent width of \MgII\ emission in \AA \\
\hline
56   & BAL\_FLAG\_VI                   &  INT32        & BAL flag from visual inspection \\
57   & BI\_CIV                         &  DOUBLE       & Balnicity index of \CIV\ trough (in ${\rm km \ s^{-1}}$) \\
58   & ERR\_BI\_CIV                    &  DOUBLE       & Error on the Balnicity index of \CIV\ trough (in ${\rm km \ s^{-1}}$) \\
59   & AI\_CIV                         &  DOUBLE       & Absorption index of \CIV\ trough (in ${\rm km \ s^{-1}}$) \\
60   & ERR\_AI\_CIV                    &  DOUBLE       & Error on the absorption index of \CIV\ trough (in ${\rm km \ s^{-1}}$) \\
61   & CHI2THROUGH                     &  DOUBLE       & $\chi^2$ of the trough from Eq. (3) in \cite{paris2012}\\
62   & NCIV\_2000                      &  INT32        & Number of distinct \CIV\ troughs of width larger than \numprint{2000}~km~s$^{-1}$ \\
63   & VMIN\_CIV\_2000                 &  DOUBLE       & Minimum velocity of the \CIV\ troughs defined in row 57 (${\rm km \ s^{-1}}$) \\
64   & VMAX\_CIV\_2000                 &  DOUBLE       & Maximum velocity of the \CIV\ troughs defined in row 57  (in ${\rm km \ s^{-1}}$) \\
65   & NCIV\_450                       &  INT32        & Number of distinct \CIV\ troughs of width larger than 450~km~s$^{-1}$ \\
66   & VMIN\_CIV\_450                  &  DOUBLE       & Minimum velocity of the \CIV\ troughs defined in row 59 (in ${\rm km \ s^{-1}}$) \\
67   & VMAX\_CIV\_450                  &  DOUBLE       & Maximum velocity of the \CIV\ troughs defined in row 59 (in ${\rm km \ s^{-1}}$) \\
68   & REW\_SIIV                       &  DOUBLE       & rest frame equivalent width of the \SiIV\ trough \\
69   & REW\_CIV                        &  DOUBLE       & rest frame equivalent width of the \CIV\ trough  \\
70   & REW\_ALIII                      &  DOUBLE       & rest frame equivalent width of the \ion{Al}{iii]} trough \\
\hline
%
%
71   & RUN\_NUMBER                     & INT32         & SDSS Imaging Run Number of photometric measurements \\
72   & PHOTO\_MJD                      & INT32         & Modified Julian Date of imaging observation \\
73   & RERUN\_NUMBER                   & STRING        & SDSS Photometric Processing Rerun Number \\
74   & COL\_NUMBER                     & INT32         & SDSS Camera Column Number (1-6) \\
75   & FIELD\_NUMBER                   & INT32         & SDSS Field Number \\
76   & OBJ\_ID                         & STRING        & SDSS Object Identification Number \\
%
%
77   & PSFFLUX                         & FLOAT[5]      & flux in the $u$,$g$,$r$,$i$ and $z$-bands (not corrected for Galactic extinction)\\
78   & IVAR\_PSFLUX                    & FLOAT[5]      & inverse variance of $u$,$g$,$r$,$i$ and $z$ fluxes \\
79   & PSFMAG                          & FLOAT[5]      & PSF magnitudes in $u$,$g$,$r$,$i$ and $z$-bands (not corrected for Galactic \\
     &                                 &               & extinction)\\
80   & ERR\_PSFMAG                     & FLOAT[5]      & error in $u$,$g$,$r$,$i$ and $z$ PSF magnitudes\\
81   & TARGET\_FLUX                    & FLOAT[5]      & TARGET flux in the $u$,$g$,$r$,$i$ and $z$-bands (not corrected for Galactic \\
     &                                 &               & extinction)\\
82   & MI                              &  FLOAT        & $M_{\rm i}\left[{\rm z = 2} \right] \left( H_0 = 67.8 {\rm km \ s^{-1} \ Mpc^{-1}}, \ \Omega _M = 0.308, \ \Omega _{\Lambda} = 0.692, \ \alpha _{\nu} = -0.5 \right)$ \\
83   & DGMI                            &  FLOAT        & $\Delta (g-i) = (g-i) - \langle (g-i) \rangle _{\rm redshift}$ (Galactic extinction corrected) \\
84   & GAL\_EXT                        & FLOAT[5]      & Galactic extinction in the 5 SDSS bands from \cite{schlegel1998} \\
85   & GAL\_EXT\_RECAL                 & FLOAT[5]      & Galactic extinction in the 5 SDSS bands from \cite{schlafly2011}\\
86   & HI\_GAL                         &    FLOAT      & ${\rm log} N_{\rm H}$ (logarithm of Galactic \HI\ column density in ${\rm cm^{-2}}$)\\
\hline
%
%
87   & VAR\_MATCHED                    &  SHORT        & Flag for variability information (0 if not available) \\
88   & VAR\_CHI2                       &  DOUBLE       & Reduced $\chi ^2$ when the light curve is fit with a constant \\
89   & VAR\_A                          &  DOUBLE       & Structure function parameter $A$ as defined \\
     &                                 &               & in \cite{palanque2011} \\
90   & VAR\_GAMMA                      &  DOUBLE       & Structure function parameter $\gamma $ as defined \\
     &                                 &               & in \cite{palanque2011} \\
\hline
%
%
91   & RASS\_COUNTS                    &  DOUBLE       & log RASS full band count rate (counts s$^{-1}$)\\
92   & RASS\_COUNTS\_SNR               &  FLOAT        & S/N of the RASS count rate \\
93   & SDSS2ROSAT\_SEP                 &  DOUBLE       & SDSS-RASS separation in arcsec \\
94   & N\_DETECTION\_XMM               & INT32         & Number of detections in XMM-Newton \\
95   & FLUX02\_12KEV\_SGL              & DOUBLE        & Total flux (0.2-12 keV) from XMM-Newton computed from the longest \\
     &                                 &               & observation (${\rm erg \ cm^{-2} \ s^{-1}}$) \\
96   & ERR\_FLUX02\_12KEV\_SGL         & DOUBLE        & Error in total flux (0.2-12 keV) from XMM-Newton \\
97   & FLUX02\_2KEV                    & DOUBLE        & Soft (0.2-2 keV) X-ray flux from XMM-Newton (${\rm erg \ cm^{-2} \ s^{-1}}$) \\
98   & ERR\_FLUX02\_2KEV               & DOUBLE        & Error in soft (0.2-2 keV) X-ray flux from XMM-Newton (${\rm erg \ cm^{-2} \ s^{-1}}$) \\
99   & FLUX45\_12KEV                   & DOUBLE        & Hard (4.5-12 keV) X-ray flux from XMM-Newton (${\rm erg \ cm^{-2} \ s^{-1}}$) \\
100  & ERR\_FLUX45\_12KEV              & DOUBLE        & Error in hard (4.5-12 keV) X-ray flux from XMM-Newton (${\rm erg \ cm^{-2} \ s^{-1}}$) \\
101  & FLUX02\_12KEV                   & DOUBLE        & Total weighted average flux (0.2-12 keV) from XMM-Newton (${\rm erg \ cm^{-2} \ s^{-1}}$) \\
102  & ERR\_FLUX02\_12KEV              & DOUBLE        & Error in total flux (0.2-12 keV) from XMM-Newton  \\
103  & LUM02\_12KEV\_SGL               & DOUBLE        & Total X-ray luminosity (0.2-12 keV) from XMM-Newton using the \\
     &                                 &               & longest observation (${\rm erg \ s^{-1}}$) using Z\_VI and $H_0 = 70 {\rm km \ s^{-1} \ Mpc^{-1}}, $ \\
     &                                 &               &  $\Omega _M = 0.3, \  \Omega _{\Lambda} = 0.7$\\
104  & LUM05\_2KEV                     & DOUBLE        & Soft (0.2-2 keV) X-ray luminosity from XMM-Newton (${\rm erg \ s^{-1}}$)  \\
     &                                 &               & using Z\_VI and $H_0 = 70 {\rm km \ s^{-1} \ Mpc^{-1}}, \ \Omega _M = 0.3, \ \Omega _{\Lambda} = 0.7$\\
105  & LUM45\_12KEV                    & DOUBLE        & Hard (4.5-12 keV) X-ray luminosity from XMM-Newton (${\rm erg \ s^{-1}}$) \\
     &                                 &               & using Z\_VI and $H_0 = 70 {\rm km \ s^{-1} \ Mpc^{-1}}, \ \Omega _M = 0.3, \ \Omega _{\Lambda} = 0.7$\\
106  & LUM2\_12KEV                     & DOUBLE        & Total X-ray luminosity (0.2-12 keV) from XMM-Newton using weighted   \\
     &                                 &               & average flux (${\rm erg \ s^{-1}}$) using Z\_VI and $H_0 = 70 {\rm km \ s^{-1} \ Mpc^{-1}}, \ \Omega _M = 0.3,$ \\
     &                                 &               & $\Omega _{\Lambda} = 0.7$\\
107  & LUMX2\_10\_UPPER                & DOUBLE        & Flag for upper limit of hard X-ray flux (in col. \#95) \\
108  & SDSS2XMM\_SEP                   & DOUBLE        & SDSS-XMM-Newton separation in arcsec \\
109  & GALEX\_MATCHED                  & SHORT         & GALEX match \\
110  & FUV                             & DOUBLE        & $fuv$ flux (GALEX) \\
111  & FUV\_IVAR                       & DOUBLE        & Inverse variance of $fuv$ flux \\
112  & NUV                             & DOUBLE        & $nuv$ flux (GALEX) \\
113  & NUV\_IVAR                       & DOUBLE        & Inverse variance of $nuv$ flux \\
114  & JMAG                            & DOUBLE        & $J$ magnitude (Vega, 2MASS) \\
115  & ERR\_JMAG                       & DOUBLE        & Error in $J$ magnitude \\
116  & JSNR                            & DOUBLE        & J-band S/N \\
117  & JRDFLAG                         & INT32         & J-band photometry flag\\ 
118  & HMAG                            & DOUBLE        & $H$ magnitude (Vega, 2MASS) \\
119  & ERR\_HMAG                       & DOUBLE        & Error in $H$ magnitude \\
120  & HSNR                            & DOUBLE        & H-band S/N \\
121  & HRDFLAG                         & INT32         & H-band photometry flag\\ 
122  & KMAG                            & DOUBLE        & $K$ magnitude (Vega, 2MASS) \\
123  & ERR\_KMAG                       & DOUBLE        & Error in $K$ magnitude \\
124  & KSNR                            & DOUBLE        & K-band S/N \\
125  & KRDFLAG                         & INT32         & K-band photometry flag\\ 
126  & SDSS2MASS\_SEP                  & DOUBLE        & SDSS-2MASS separation in arcsec \\
127  & W1MAG                           & DOUBLE        & $w1$ magnitude (Vega, WISE) \\
128  & ERR\_W1MAG                      & DOUBLE        & Error in $w1$ magnitude \\
129  & W1SNR                           & DOUBLE        & S/N in w1 band    \\
130  & W1CHI2                          & DOUBLE        & $\chi^2$ in w1 band \\
131  & W2MAG                           & DOUBLE        & $w2$ magnitude (Vega, WISE) \\
132  & ERR\_W2MAG                      & DOUBLE        & Error in $w2$ magnitude \\
133  & W2SNR                           & DOUBLE        & S/N in w2 band    \\
134  & W2CHI2                          & DOUBLE        & $\chi^2$ in w2 band \\
135  & W3MAG                           & DOUBLE        & $w3$ magnitude (Vega, WISE) \\
136  & ERR\_W3MAG                      & DOUBLE        & Error in $w3$ magnitude \\
137  & W3SNR                           & DOUBLE        & S/N in w3 band    \\
138  & W3CHI2                          & DOUBLE        & $\chi^2$ in w3 band \\
139  & W4MAG                           & DOUBLE        & $w4$ magnitude  (Vega, WISE) \\
140  & ERR\_W4MAG                      & DOUBLE        & Error in $w4$ magnitude \\
141  & W4SNR                           & DOUBLE        & S/N in w4 band    \\
142  & W4CHI2                          & DOUBLE        & $\chi^2$ in w4 band \\
143  & CC\_FLAGS                       & STRING        & WISE contamination and confusion flag \\
144  & PH\_FLAG                        & STRING        & WISE photometric quality flag \\
145  & SDSS2WISE\_SEP                  & DOUBLE        & SDSS-WISE separation in arcsec \\
146  & UKIDSS\_MATCHED                 & SHORT         & UKIDSS Matched \\
147  & YFLUX                           & DOUBLE        & Y-band flux density from UKIDSS (in ${\rm W \ m^{-2} \ Hz^{-1}}$) \\
148  & YFLUX\_ERR                      & DOUBLE        & Error in Y-band density flux from UKIDSS (in ${\rm W \ m^{-2} \ Hz^{-1}}$) \\
149  & JFLUX                           & DOUBLE        & J-band flux density from UKIDSS (in ${\rm W \ m^{-2} \ Hz^{-1}}$)\\
150  & JFLUX\_ERR                      & DOUBLE        & Error in J-band flux density from UKIDSS (in ${\rm W \ m^{-2} \ Hz^{-1}}$) \\
151  & HFLUX                           & DOUBLE        & H-band flux density from UKIDSS (in ${\rm W \ m^{-2} \ Hz^{-1}}$)\\
152  & HFLUX\_ERR                      & DOUBLE        & Error in H-band flux density from UKIDSS (in ${\rm W \ m^{-2} \ Hz^{-1}}$)\\
153  & KFLUX                           & DOUBLE        & K-band flux density from UKIDSS (in ${\rm W \ m^{-2} \ Hz^{-1}}$)\\
154  & KFLUX\_ERR                      & DOUBLE        & Error in K-band flux density from UKIDSS  (in ${\rm W \ m^{-2} \ Hz^{-1}}$) \\
155  & FIRST\_MATCHED                  & INT           & FIRST matched \\
156  & FIRST\_FLUX                     & DOUBLE        & FIRST peak flux density at 20 cm expressed in mJy \\
157  & FIRST\_SNR                      & DOUBLE        & S/N of the FIRST flux density \\
158  & SDSS2FIRST\_SEP                 & DOUBLE        & SDSS-FIRST separation in arcsec \\
\hline
\multicolumn{4}{l}{$^a$ All magnitudes are PSF magnitudes}
\end{longtable}
}

\par\noindent
Notes on the catalog columns:\\
\smallskip
\noindent
1. The DR12 object designation, given by the format \hbox{SDSS Jhhmmss.ss+ddmmss.s}; only the final 18
characters  are listed in the catalog (i.e., the character string \hbox{"SDSS J"}  is dropped).
The coordinates in the object name follow IAU convention and are truncated, not rounded.

\noindent
2-3. The J2000 coordinates (Right Ascension and Declination) in decimal degrees.  
The astrometry is from SDSS-DR12 \citep{DR12}.

\noindent
4.  The 64-bit integer that uniquely describes the objects
that are listed in the SDSS (photometric and spectroscopic) catalogs ({\tt THING\_ID}).

\noindent
5-7. Information about the spectroscopic observation (Spectroscopic plate number, 
Modified Julian Date, and spectroscopic fiber number) used to
determine the characteristics of the spectrum.
These three numbers are unique for each spectrum, and
can be used to retrieve the digital spectra from the public SDSS database.
When an object has been observed more than once, we selected the best quality spectrum as 
defined by the SDSS pipeline \citep{bolton2012}, i.e. with {\tt SPECPRIMARY}~=~1.

\noindent
8. Redshift from the visual inspection, {\tt Z\_VI}.

\noindent
9. Redshift from the BOSS pipeline \citep{bolton2012}.

\noindent
10. Error on the BOSS pipeline redshift estimate.

\noindent
11. ZWARNING flag from the pipeline. ZWARNING~$>$~0 indicates uncertain results in 
the redshift-fitting code \citep{bolton2012}.

\noindent
12.  Automatic redshift estimate using a linear combination of four principal components \citep[see Section 4 of ][ for details]{paris2012}.
When the velocity difference between the automatic PCA and visual inspection redshift estimates is larger than 
\numprint{5000}~${\rm km \ s^{-1}}$, this PCA redshift is set to $-1$.

\noindent
13. Error on the automatic PCA redshift estimate.
If the PCA redshift is set to $-1$, the associated error is also set to $-1$.

\noindent
14. Estimator of the PCA continuum quality (between 0 and 1) as given in Eq. 11 of \cite{paris2011}.

\noindent
15-17. Redshifts measured from the \CIV , the \CIII\ complex and the \MgII\ emission lines 
from a linear combination of five principal components \citep[see ][]{paris2012}. 
The line redshift is estimated using the position of the maximum of each emission line, contrary to {\tt Z\_PCA} (column \#12) which
is a global estimate using all the information available in a given spectrum.

\noindent
18. Morphological information: objects classified as a point source by the SDSS photometric pipeline have {\tt SDSS\_MORPHO}~=~0 while extended quasars have {\tt SDSS\_MORPHO}~=~1.
The vast majority of the quasars included in the present catalog are unresolved ({\tt SDSS\_MORPHO}~=~0) as this is a requirement of the main quasar target selection \citep{ross2012}. 

\noindent
19-22.
The main target selection information is tracked with the {\tt BOSS\_TARGET1} flag bits \citep[19; see Table 2 in ][ for a full description]{ross2012}.
Ancillary program target selection is tracked 
with the {\tt ANCILLARY\_TARGET1} (20) and {\tt ANCILLARY\_TARGET2} (21) flag bits.
The bit values and the corresponding program names are listed in \cite{dawson2013}, \cite{DR12}, and Appendix B of this paper. 
The SEQUELS program targeted different classes of objects (quasars, LRG, galaxy clusters).
All the SEQUELS targets are identified by the {\tt ANCILLARY\_TARGET2} bit 53, and the details of each target class is provided through
the {\tt EBOSS\_TARGET0} flag (Col. \#22; see \Tab{VIres_detail}).

\noindent
23-26. If a quasar in DR12Q was observed more than once during the survey, the number of additional SDSS-III/BOSS spectra is given by {\tt NSPEC\_BOSS} (Col. \#23).
The associated plate ({\tt PLATE\_DUPLICATE}), MJD ({\tt MJD\_DUPLICATE}) and fiber ({\tt FIBERID\_DUPLICATE}) numbers are given in Col. \#24, 25 and 26 respectively.
If a quasar was observed N times in total, the best spectrum is identified in Col. \#5-7, the corresponding {\tt NSPEC\_BOSS} is N-1, and the first N-1 columns of {\tt PLATE\_DUPLICATE},
{\tt MJD\_DUPLICATE} and {\tt FIBERID\_DUPLICATE} are filled with relevant information. Remaining columns are set to -1.

\noindent
27. A quasar previously known from the SDSS-DR7 quasar catalog  has an entry equal to 1, and 0 otherwise.
During Year 1 and 2, most SDSS-DR7 quasars with $z \geq 2.15$ were re-observed.
After Year 2, SDSS-DR7 quasars with $z \geq 1.8$ were systematically re-observed.

\noindent
28-30. Spectroscopic plate number, Modified Julian Date, and spectroscopic fiber number
in SDSS-DR7.

\noindent
31. {\tt Uniform }flag. See \Sec{uniform}.

\noindent
32. Spectral index $\alpha _{\nu}$.
The continuum is approximated by a power-law, $f_{cont} \propto \nu ^{\alpha _{\nu}}$, and is computed in emission line free regions: \numprint{1450}-\numprint{1500} \AA , \numprint{1700}-\numprint{1850} \AA\ and \numprint{1950}-\numprint{2750}\AA\ in the quasar rest frame.

\noindent
33. Median signal-to-noise ratio per pixel computed over the whole spectrum for the best spectrum as defined by the SDSS pipeline, i.e., with {\rm SPECPRIMARY}~=~1 \citep{bolton2012}.

\noindent
34. Median signal-to-noise ratio per pixel computed over the whole spectrum of quasars observed multiple times.
If there is no multiple spectroscopic observation available, the S/N value is set to $-1$.

\noindent
35-37.  Median signal-to-noise ratio per pixel computed over the windows \numprint{1650}-\numprint{1750} \AA\ (Col. \#35),
\numprint{2950}-\numprint{3050} \AA\ (Col. \#36) and \numprint{2950}-\numprint{3050} \AA\ (Col. \#37) in the quasar rest frame. If the wavelength 
range is not covered by the BOSS spectrum, the value is set to $-1$.

\noindent
38-41. FWHM (${\rm km \ s^{-1}}$), blue and red half widths at half-maximum (HWHM; the sum of the latter two equals FWHM), 
and amplitude \citep[in units of the median rms pixel noise, see Section~4 of ][]{paris2012} of the \CIV\  emission line. 
If the emission line is not in the spectrum, the red and blue HWHM and the FWHM are set to -1.

\noindent
42-43. Rest frame equivalent width and corresponding uncertainty in \AA\ of the \CIV\ emission line. If the emission line is not in the 
spectrum, these quantities are set to -1.

\noindent
44-47. 
Same as 38-41 for the  \ion{C}{iii]} emission complex.
It is well known that \ion{C}{iii]}${\rm \lambda \lambda }$1909 is blended with \ion{Si}{iii]}${\rm \lambda \lambda }$1892
and to a lesser extent with \ion{Al}{iii]}${\rm \lambda \lambda }$1857. We do not attempt
to deblend these lines. Therefore the redshift and red side of the HWHM derived for this blend correspond to
\ion{C}{iii]}${\rm \lambda \lambda }$1909. The blue side of the HWHM is obviously affected by the blend.

\noindent
48-49. Rest frame equivalent width and corresponding uncertainty in \AA\ of the \CIII\ emission complex. 

\noindent
50-53.  Same as 38-41 for the \MgII\ emission line. 

\noindent
54-55. Rest frame equivalent width and corresponding uncertainty in \AA\ of the \MgII\ emission line. We do not correct for the neighboring \ion{Fe}{ii} emission.\\
Note that \cite{albareti2015} released the output of \ion{[O}{iii]}${\rm \lambda}$4960,5008\AA\ emission-line fitting for a subset of the present catalog. 

\noindent
56. BAL flag from the visual inspection, {\tt BAL\_FLAG\_VI}. 
If a BAL feature was identified in the course of the visual inspection, {\tt BAL\_FLAG\_VI} is set to 1, the flag is set to 0 otherwise.
BAL quasars are flagged during the visual inspection at any redshift and whenever a BAL feature is seen
in the spectrum (not only for \ion{C}{iv} absorption troughs).

\noindent
57-58. Balnicity index \citep[BI; ][]{weymann1991} for \CIV\ troughs, and their errors, expressed in ${\rm km \ s^{-1}}$. See definition in \Sec{bal}.
The Balnicity index is measured for quasars with $z > 1.57$ only, so that the trough enters into the BOSS wavelength region.
If the BAL flag from the visual inspection is set to 1 and the BI is equal to 0, this means either that there is no 
\CIV\ trough (but a trough is seen in another transition) or that the trough seen during the visual inspection does 
not meet the formal requirement of the BAL definition. 
In cases with poor fits to the continuum, the balnicity index and its error are set to $-1$.

\noindent
59-60. Absorption index, and its error, for \CIV\ troughs expressed in ${\rm km \ s^{-1}}$. See definition in \Sec{bal}.
In cases with a poor continuum fit, the absorption index and its error are set to $-1$.

\noindent
61. Following \cite{trump2006}, we calculate the reduced $\chi^2$ which we call $\chi^2 _{{\rm trough}}$ for each \CIV\ trough
from \Eq{chi2_trough}.
We require that troughs have $\chi^2_{\rm trough}$~$>$~10 to be considered as
true troughs.

\noindent
62. Number of \CIV\ troughs of width larger than \numprint{2000}~km~s$^{-1}$.

\noindent
63-64. Limits of the velocity range in which 
\CIV\ troughs of width larger than \numprint{2000}~km~s$^{-1}$ 
and reaching at least 10\% below the continuum 
are to be found.  Velocities are positive bluewards and the zero of the
scale is at {\tt Z\_VI}.
So if there are multiple troughs, this value demarcates the range from the first to the last trough.

\noindent 
65. Number of troughs of width larger than 450~km~s$^{-1}$.

\noindent
66-67. Same as 63-64 for \CIV\ troughs of width larger than 450~km~s$^{-1}$.

\noindent
68-70. Rest frame equivalent width in \AA\ of \SiIV, \CIV\ and \AlIII\  troughs detected in BAL quasars
with {\tt BI\_CIV} $>$ 500 ${\rm km \ s^ {-1}}$ and {\tt SNR\_1700} $>$~5 (see Col. \#35).
They are set to 0 otherwise or in cases where no trough is detected and to $-1$ if the continuum is not reliable.

\noindent
71-72. The SDSS Imaging Run number ({\tt RUN\_NUMBER}) and the Modified Julian Date (MJD) of the
photometric observation used in the catalog ({\tt PHOTO\_MJD}).

\noindent
73-76. Additional SDSS processing information: the
photometric processing rerun number ({\tt RERUN\_NUMBER}); the camera column (1--6) containing
the image of the object ({\tt COL\_NUMBER}), the field number of the run containing the object ({\tt FIELD\_NUMBER}),
and the object identification number
\citep[{\tt OBJ\_ID}, see][for descriptions of these parameters]{stoughton2002}.

\noindent
77-78. DR12 PSF fluxes, expressed in nanomaggies\footnote{See https://www.sdss3.org/dr8/algorithms/magnitudes.php\#nmgy}, and inverse variances (not corrected for Galactic extinction) in the five SDSS filters. 

\noindent
79-80. DR12 PSF AB magnitudes \citep{oke1983} and errors (not corrected for Galactic extinction) in the five SDSS filters \citep{lupton1999}.
These magnitudes are Asinh magnitudes as defined in \cite{lupton1999}.

\noindent
81. DR8 PSF fluxes (not corrected for Galactic extinction), expressed in nanommagies, in the five SDSS filters. 
This photometry is the one that was used for the main quasar target selection \citep{ross2012}.

\noindent
82. The absolute magnitude in the $i$ band at $z=2$ calculated 
using a power-law (frequency)
continuum index of~$-0.5$.
The K-correction is computed using Table~4 from \cite{richards2006}.
We use the SDSS primary photometry to compute this value.

\noindent
83. The $\Delta (g-i)$ color, which is the difference in the Galactic
extinction corrected $(g-i)$ for the quasar and that of the mean of the
quasars at that redshift .  If $\Delta (g-i)$ is not defined for the quasar,
which occurs for objects at either \hbox{$z < 0.12$} or \hbox{$z > 5.12$};
the column will contain~``$-9999$".

\noindent
84. Galactic extinction in the five SDSS bands based on the maps of
\cite{schlegel1998}. The quasar target selection was done using these maps.

\noindent
85. Galactic extinction in the five SDSS bands based on \cite{schlafly2011}.

\noindent
86. The logarithm of the Galactic neutral hydrogen column density along the
line of sight to the quasar. These values were
estimated via interpolation of the 21-cm data from \cite{stark1992},
using the COLDEN software provided by the {\it Chandra} X-ray Center.
Errors associated with the interpolation are expected to
be typically less than $\approx 1\times 10^{20}$~cm$^{-2}$ 
\citep[e.g., see \S5 of][]{elvis1994b}.

\noindent
87. Flag for variability information. If no information is available,
{\tt VAR\_MATCHED} is set to 0.
When enough photometric epochs are available to build a lightcurve,
{\tt VAR\_MATCHED} is equal to the number of photometric epochs used.
For PTF-selected objects ({\tt EBOSS\_TARGET0}, bit 11), both PTF and SDSS photometry were used.
For objects lying in the stripe 82 region, only SDSS photometry was used.

\noindent
88. Light curves were fit by a constant flux \citep[see Sec. 2.3 in ][]{palanque2011}.
The reduced $\chi ^2$ of this fit is provided in {\tt VAR\_CHI2}.

\noindent
89-90. The best fit values of the structure function parameters $A$ and $\gamma$, as defined in \Eq{str_func}, are reported in Col. 89 ({\tt VAR\_A}) and 90 ({\tt VAR\_GAMMA}), respectively.

\noindent
91. The logarithm of the vignetting-corrected count rate (photons s$^{-1}$)
in the broad energy band \hbox{(0.1--2.4 keV)} from the
{\it ROSAT} All-Sky Survey Faint Source Catalog \citep{voges2000} and the
{\it ROSAT} All-Sky Survey Bright Source Catalog \citep{voges1999}.
The matching radius was set to 30\arcsec \ (see \Sec{rosat}).

\noindent
92. The S/N of the {\it ROSAT} measurement.

\noindent
93. Angular separation between the SDSS and {\it ROSAT} All-Sky Survey
locations (in arcseconds).

\noindent
94. Number of XMM-Newton matches in a 5\arcsec\ radius around the SDSS-DR12 quasar positions.

\noindent
95-96. Total X-ray flux (0.2--12 keV) from the three XMM-Newton CCDs (MOS1, MOS2 and PN), expressed in ${\rm erg \ cm^{-2} \ s^{-1}}$, and its error. In the case of multiple XMM-Newton observations, only the longest
exposure was used to compute the reported flux.

\noindent
97-98. Soft X-ray flux (0.2--2 keV) from XMM-Newton matching, expressed in ${\rm erg \ cm^{-2} \ s^{-1}}$, and its error. 
In the case of multiple observations, the values reported here are the weighted average of all the XMM-Newton detections in this band.

\noindent
99-100. Hard X-ray flux (4.5--12 keV) from XMM-Newton matching, expressed in ${\rm erg \ cm^{-2} \ s^{-1}}$, and its error.
In the case of multiple observations, the values reported here are the weighted average of all the XMM-Newton detections in this band.

\noindent
101-102 Total X-ray flux (0.2--12 keV) from XMM-Newton matching, expressed in ${\rm erg \ cm^{-2} \ s^{-1}}$, and its error.
In the case of multiple observations, the values reported here are the weighted average of all the XMM-Newton detections in this band.
For single observations, the total X-ray flux reported in Col. 101 ({\tt FLUX02\_12KEV}) is equal to the one reported in Col. 95 ({\tt FLUX02\_12KEV\_SGL}).

\noindent
103. Total X-ray luminosity (0.2--12 keV) derived from the flux computed in Col. \#95, expressed in ${\rm erg \ s^{-1}}$. 
This value is computed using the visual inspection redshift ({\tt Z\_VI})
and is not absorption corrected.

\noindent
104. X-ray luminosity in the 0.5--2 keV band of XMM-Newton, expressed in ${\rm erg \ s^{-1}}$. 
This value is computed using the visual inspection redshift ({\tt Z\_VI}),  ${\rm H_0 = 70 \ km s^{-1} Mpc^{-1}, \ \Omega _m = 0.3, \ \Omega _{\Lambda} = 0.7}$ and is not absorption corrected.
 
\noindent
105. X-ray luminosity in the 4.5--12 keV band of XMM-Newton, expressed in ${\rm erg \ s^{-1}}$. 
This value is computed using the visual inspection redshift ({\tt Z\_VI}),   ${\rm H_0 = 70 \ km s^{-1} Mpc^{-1}, \ \Omega _m = 0.3, \ \Omega _{\Lambda} = 0.7}$ and is not absorption corrected.

\noindent
106. Total X-ray luminosity (0.2--12 keV) using the flux value reported in Col. 101.
This value is computed using the visual inspection redshift ({\tt Z\_VI}),   ${\rm H_0 = 70 \ km s^{-1} Mpc^{-1}, \ \Omega _m = 0.3, \ \Omega _{\Lambda} = 0.7}$  and is not absorption corrected.

\noindent
107. In the case of an unreliable detection or no detection 
in the 2--10 keV band the flux reported in Col. 101 is an upper limit. 
In that case, the {\tt LUMX2\_10\_UPPER} flag listed in this column is set to $-1$.

\noindent
108. Angular separation between the XMM-Newton and SDSS-DR12 locations, expressed in arcsec.

\noindent
109. If a SDSS-DR12 quasar matches with GALEX photometring, {\tt GALEX\_MATCHED} is set to 1, 0 if not.

\noindent
110-113. UV fluxes and inverse variances from GALEX, aperture-photometered from the original GALEX images in the 
two bands FUV and NUV. The fluxes are expressed in nanomaggies.

\noindent
114-115. The $J$ magnitude and error from the Two Micron All Sky Survey
All-Sky Data Release Point Source Catalog \citep{cutri2003} using
a matching radius of ~2.0\arcsec\ (see \Sec{2mass}).  A non-detection by 2MASS is indicated by a "0.000" in these columns.  
The 2MASS measurements are in Vega, not AB, magnitudes.  

\noindent
116-117. Signal-to-noise ratio in the $J$ band and corresponding 2MASS {\tt jr\_d} flag that gives the meaning of the peculiar values of the magnitude and its error\footnote{see http://www.ipac.caltech.edu/2mass/releases/allsky/doc/ explsup.html}.

\noindent
118-121. Same as 114-117 for the $H$-band.

\noindent
122-125. Same as 114-117 for the $K$-band.

\noindent
126. Angular separation between the SDSS-DR12 and 2MASS positions (in arcsec).

\noindent
127-128. The $w1$  magnitude and error from the Wide-field Infrared Survey Explorer
\citep[WISE;][]{wright2010} AllWISE Data Release Point Source Catalog  using a matching radius of 2\arcsec.

\noindent
129-130 Signal-to-noise ratio and $\chi^2$ in the WISE $w1$ band.

\noindent
131-134. Same as 127-130 for the $w2$-band.

\noindent
135-138. Same as 127-130 for the $w3$-band.

\noindent
139-142. Same as 127-130 for the $w4$-band.

\noindent
143. WISE contamination and confusion flag.

\noindent
144. WISE photometric quality flag.

\noindent
145. Angular separation between SDSS-DR12 and WISE positions (in arcsec).

\noindent
146. If a SDSS-DR12 quasar matches UKIDSS aperture-photometering data, {\tt UKIDSS\_MATCHED} is set to 1, it is set to 0 if not.

\noindent
147-154. Flux density and error from UKIDSS, aperture-photometered from the original UKIDSS images in the four bands Y (Col. \#147-148), J (Col. \#149-150), 
H (Col. \#151-152) and K (Col. \#153-154). The fluxes and errors are expressed in ${\rm W \ m^{-2} \ Hz^{-1}}$.

\noindent
155. If there is a source in the FIRST radio catalog (version March 2014) within 2.0\arcsec\
of the quasar position, the {\tt FIRST\_MATCHED} flag provided in this column is set to 1, 0 if not. If the quasar lies outside of the FIRST footprint, it is set to -1.

\noindent
156. This column contains the FIRST peak flux density, expressed in mJy.

\noindent
157. The signal-to-noise ratio of the FIRST source whose flux is given in Col. \#156.

\noindent
158. Angular separation between the SDSS-DR12 and FIRST positions (in arcsec).

%
\section{Supplementary lists}	
\label{s:suplist}

\begin{figure}[htbp]
 \centering{\includegraphics[width=75mm]{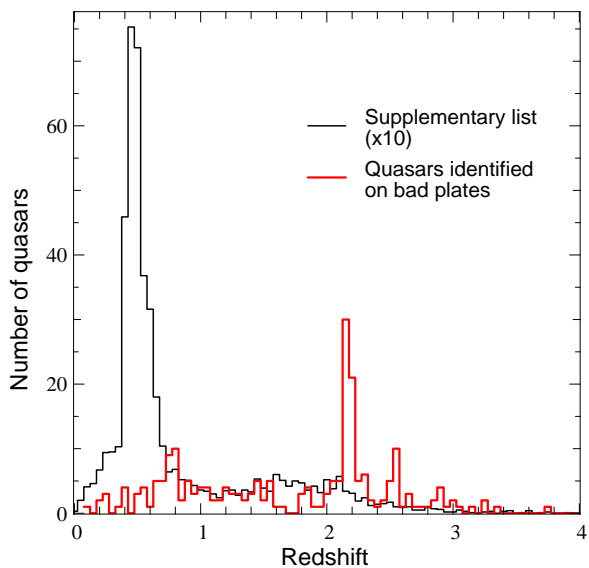}}
 \caption{Redshift distribution of quasars serendipitously identified quasars (black histogram) and 
 identified on bad plates (red histogram). The number of objects for the black histogram
 is divided by 10 for clarity.}
 \label{fig:zdistri_sup}
\end{figure}

As explained in \Sec{construction}, visual inspection is systematically performed for all the quasar targets of
the SDSS-III/BOSS survey. Some galaxy targets are also included in the superset from which we derive DR12Q
based on their SDSS pipeline identification.
The latter depends on the SDSS pipeline version. Most identifications remain unchanged between the two versions.
Nevertheless a few galaxy targets that met the requirements to be part of the quasar catalog superset 
when their spectra were reduced by a former version of the pipeline
no longer comply with these requirements when the most recent version of the pipeline is used. 
Firmly identified quasars that drop out of DR12Q for this reason are included in
the supplementary list.
The visual inspection strategy for galaxy targets has also evolved over the course of the survey. Quasars identified
during these tests are also part of the supplemental list.
This first list contains a total of \numprint{4841} quasars. The redshift distribution of this sample is shown in
\Fig{zdistri_sup} (black histogram).
The spectra of all those quasars are available though the SDSS website.\\

In addition to this first supplementary list, we provide a list of quasars observed during the commissioning phase
of the SDSS-III/BOSS spectrograph. Those objects were observed between MJD~=~55103 and MJD~=~55177 and were not re-observed
later in the survey. This second supplementary list contains \numprint{291} quasars whose redshift distribution is also
shown in \Fig{zdistri_sup} (red histogram). 
The spectra of those quasars are not part of the official DR12 \citep{DR12} and are not available though the standard
SDSS website. We provide an archival file that contains those \numprint{291} spectra.\\

\begin{table*}
\centering                        
\begin{tabular}{c l c l}        
\hline\hline
Column & Name & Format & Description\\
\hline\hline
1    & SDSS\_NAME                      &  STRING      & SDSS-DR12 designation  hhmmss.ss+ddmmss.s (J2000)\\
2    & RA                              &  DOUBLE      & Right Ascension in decimal degrees (J2000)\\
3    & DEC                             & DOUBLE       & Declination in decimal degrees (J2000)\\
4    & THING\_ID                       &  INT32       & Thing\_ID \\
5    & PLATE                           & INT32        & Spectroscopic Plate number \\
6    & MJD                             & INT32        & Spectroscopic MJD \\
7    & FIBERID                         & INT32        & Spectroscopic Fiber number \\
\hline
8    & Z\_VI                           &  DOUBLE      & Redshift from visual inspection \\
9    & Z\_PIPE                         &  DOUBLE      & Redshift from BOSS pipeline \\
10   & ERR\_ZPIPE                      &  DOUBLE      & Error on BOSS pipeline redshift \\
11   & ZWARNING                        & INT32        & ZWARNING flag  \\
\hline
12   & SDSS\_MORPHO                    &   INT32      & SDSS morphology flag 0 = point source 1 = extended \\
13   & BOSS\_TARGET1                   &  INT64       & BOSS target flag for main survey  \\
14   & ANCILLARY\_TARGET1              &   INT64      & BOSS target flag for ancillary programs \\
15   & ANCILLARY\_TARGET2              &   INT64      & BOSS target flag for ancillary programs  \\
16   & EBOSS\_TARGET0                  &   INT64      & Target selection flag for eBOSS pilot survey \\
\hline
17   & BAL\_FLAG\_VI                   & SHORT        & BAL flag from visual inspection \\
\hline
18   & OBJ\_ID                         & STRING       & SDSS Object Identification Number \\
17   & PSFFLUX                         & FLOAT[5]     & flux in the $u$,$g$,$r$,$i$ and $z$-bands (not corrected for Galactic extinction)\\
18   & IVAR\_PSFLUX                    & FLOAT[5]     & inverse variance of $u$,$g$,$r$,$i$ and $z$ fluxes \\
19   & PSFMAG                          & FLOAT[5]     & PSF magnitudes in $u$,$g$,$r$,$i$ and $z$-bands (not corrected for Galactic \\
     &                                 &              & extinction)\\
20   & ERR\_PSFMAG                     & FLOAT[5]     & error in $u$,$g$,$r$,$i$ and $z$ PSF magnitudes\\
21   & GAL\_EXT                        & FLOAT[5]     & Galactic extinction in the 5 SDSS bands \citep[from ][]{schlegel1998} \\
\hline
\hline
\end{tabular}
\caption{Format of the binary FITS files containing the two supplementary lists.}
\label{t:supplementary}
\end{table*}

The format of these two supplemental lists is detailed in \Tab{supplementary}.\\

Finally, these supplementary lists do not contain the spectra of all
the AGNs available in the SDSS spectroscopic sample. In particular, there are objects targeted
by the SDSS-III/BOSS galaxy survey that are spectroscopically classified as galaxies, but which
show strong broad \ion{Mg}{ii} in emission \citep{roig2014,pattarakijwanich2014}. These objects
could not be selected as part of the superset described in \Sec{construction} because of their
classification by the SDSS pipeline.

%
\section{Conclusion}	
\label{s:conclusion}

We have presented the final quasar catalog of the SDSS-III/BOSS survey resulting from five years of observations.
The catalog, which we call ``DR12Q'', contains \numprint{297301} quasars, \numprint{184101} of which have $z > 2.15$.
We provide robust identification from visual inspection and refined redshift measurements based on the result of
a principal component analysis of the spectra.
The present catalog contains about 80\% more quasars than our previous release \citep{paris2014}.
As part of DR12Q, we also provide a catalog of \numprint{29579} BAL quasars and their properties.
Spectroscopic line measurements are provided together with multi-wavelength matching with GALEX, 2MASS, UKIDSS, WISE, 
FIRST, RASS and XMM-Newton observations.

Observations for the fourth stage of the SDSS (SDSS-IV) started in July 2014. As part of the extended Baryonic Oscillation Spectroscopic
Survey \citep[eBOSS; ][]{dawson2015}, about half a million new quasars will be observed \citep{myers2015}.
In keeping with the growth rate of previous quasar surveys, SDSS-IV/eBOSS will expand the number of {\em uniformly targeted} quasars
at redshifts $0.9 \leq z \leq 2.2$ by more than an order of magnitude.
The first spectroscopic release of SDSS-IV/eBOSS is expected to occur during the summer 2017.

\begin{acknowledgements}
 I.P. was supported by PRIN INAF 2012 "The X-Shooter sample of 100 quasar spectra at z-3.5: Digging into cosmology and
 galaxy evolution with quasar absorption lines".
 This work has been carried out thanks to the support of the A*MIDEX project (ANR- 11-IDEX-0001-02) funded by the “Investissements d’Avenir” French Government program, managed by the French National Research Agency (ANR).
The French Participation Group to SDSS-III was supported by the Agence Nationale de la Recherche under contracts ANR-08-BLAN-0222
and ANR-12-BS05-0015.
A.D.M was partially supported by NASA ADAP award NNX12AE38G and by NSF awards 1211112 and 1515404.\\
W.N.B. was supported by NSF grant AST-1516784.\\
I.P. thanks S. Twain, D. Hoff, D. Dintei and Brian A.J. Richardson for their inspiring contribution to this work.\\
Funding for SDSS-III has been provided by the Alfred P. Sloan Foundation, the Participating Institutions, the National Science Foundation, and the U.S. Department of Energy Office of Science. The SDSS-III web site is http://www.sdss3.org/.
SDSS-III is managed by the Astrophysical Research Consortium for the Participating Institutions of the SDSS-III Collaboration including the University of Arizona, the Brazilian Participation Group, Brookhaven National Laboratory, Carnegie Mellon University, University of Florida, the French Participation Group, the German Participation Group, Harvard University, the Instituto de Astrofisica de Canarias, the Michigan State/Notre Dame/JINA Participation Group, Johns Hopkins University, Lawrence Berkeley National Laboratory, Max Planck Institute for Astrophysics, Max Planck Institute for Extraterrestrial Physics, New Mexico State University, New York University, Ohio State University, Pennsylvania State University, University of Portsmouth, Princeton University, the Spanish Participation Group, University of Tokyo, University of Utah, Vanderbilt University, University of Virginia, University of Washington, and Yale University.\\
AllWISE makes use of data from WISE, which is a joint project of the University of California, Los Angeles, and the Jet Propulsion Laboratory/California Institute of Technology, and NEOWISE, which is a project of the Jet Propulsion Laboratory/California Institute of Technology. WISE and NEOWISE are funded by the National Aeronautics and Space Administration.\\
This research used resources of the National Energy Research Scientific Computing Center, a DOE Office of Science User Facility supported by the Office of Science of the U.S. Department of Energy under Contract No. DE-AC02-05CH11231.
\end{acknowledgements}

\bibliographystyle{aa}
\bibliography{DR12Q}

\onecolumn
\begin{appendix}
\section{List of target selection flags}
\label{ap:QTSflag}

\longtab[1]{
\begin{longtable}{l r r r r r r r r}
\caption{\label{t:VIres_detail} Detailed result of visual inspection of DR12 quasar targets for each target selection bit.}\\
\hline
\hline
Selection & Maskbits & \# Objects & \# QSO & \# QSO $z>2.15$ & \# STAR & \# GALAXY & \# ? & \# BAD \\
\hline
\endfirsthead
\caption{Continued.} \\
\hline
Selection & Maskbits & \# Objects & \# QSO & \# QSO $z>2.15$ & \# STAR & \# GALAXY & \# ? & \# BAD \\
\hline
\endhead
\hline
\endfoot
\hline
\endlastfoot
BOSS\_TARGET1 &  & & & & & & & \\ 
\hline
QSO\_CORE & 10 & 3410 & 1335 & 1070 & 1970 & 67 & 30 & 8 \\ 
QSO\_BONUS & 11 & 4212 & 773 & 455 & 3311 & 87 & 31 & 10 \\ 
QSO\_KNOWN\_MIDZ & 12 & 20249 & 20121 & 19949 & 37 & 2 & 44 & 45 \\ 
QSO\_KNOWN\_LOHIZ & 13 & 57 & 57 & 1 & 0 & 0 & 0 & 0 \\ 
QSO\_NN & 14 & 212531 & 143118 & 114815 & 64981 & 1998 & 1173 & 1261 \\ 
QSO\_UKIDSS & 15 & 47 & 27 & 22 & 18 & 2 & 0 & 0 \\ 
QSO\_LIKE\_COADD & 16 & 1326 & 314 & 228 & 889 & 53 & 52 & 18 \\ 
QSO\_LIKE & 17 & 113240 & 69037 & 47675 & 40069 & 2127 & 1044 & 963 \\ 
QSO\_FIRST\_BOSS & 18 & 8607 & 6879 & 4655 & 745 & 356 & 533 & 94 \\ 
QSO\_KDE & 19 & 238770 & 145345 & 111257 & 87802 & 2681 & 1538 & 1404 \\ 
QSO\_CORE\_MAIN & 40 & 187229 & 122692 & 101556 & 59745 & 2008 & 1448 & 1336 \\ 
QSO\_BONUS\_MAIN & 41 & 394278 & 218871 & 164897 & 158618 & 9195 & 4217 & 3377 \\ 
QSO\_CORE\_ED & 42 & 34453 & 23083 & 19578 & 10491 & 287 & 281 & 311 \\ 
QSO\_CORE\_LIKE & 43 & 36998 & 27495 & 20008 & 8485 & 480 & 253 & 285 \\ 
QSO\_KNOWN\_SUPPZ & 44 & 57 & 57 & 1 & 0 & 0 & 0 & 0 \\ 
\hline
ANCILLARY\_TARGET1 &  & & & & & & & \\ 
\hline
BLAZGVAR & 6 & 2 & 1 & 0 & 0 & 0 & 1  & 0 \\ 
BLAZR & 7 & 6 & 2 & 0 & 0 & 4 & 0  & 0 \\ 
BLAZXR & 8 & 610 & 203 & 12 & 36 & 333 & 36  & 2 \\ 
BLAZXRSAL & 9 & 3 & 3 & 1 & 0 & 0 & 0  & 0 \\ 
BLAZXRVAR & 10 & 1 & 0 & 0 & 0 & 0 & 1  & 0 \\ 
XMMBRIGHT & 11 & 623 & 489 & 29 & 15 & 114 & 5  & 0 \\ 
XMMGRIZ & 12 & 97 & 28 & 14 & 30 & 24 & 14  & 1 \\ 
XMMHR & 13 & 748 & 242 & 22 & 33 & 446 & 24  & 3 \\ 
XMMRED & 14 & 520 & 95 & 6 & 65 & 354 & 2  & 4 \\ 
FBQSBAL & 15 & 12 & 11 & 4 & 0 & 0 & 1  & 0 \\ 
LBQSBAL & 16 & 6 & 6 & 0 & 0 & 0 & 0  & 0 \\ 
ODDBAL & 17 & 32 & 24 & 8 & 1 & 1 & 6  & 0 \\ 
OTBAL & 18 & 15 & 2 & 0 & 0 & 0 & 13  & 0 \\ 
PREVBAL & 19 & 15 & 15 & 5 & 0 & 0 & 0  & 0 \\ 
VARBAL & 20 & 1470 & 1460 & 577 & 0 & 0 & 5  & 5 \\ 
QSO\_AAL & 22 & 476 & 472 & 2 & 1 & 1 & 1  & 1 \\ 
QSO\_AALS & 23 & 852 & 846 & 54 & 0 & 1 & 1  & 4 \\ 
QSO\_IAL & 24 & 278 & 277 & 3 & 0 & 0 & 1  & 0 \\ 
QSO\_RADIO & 25 & 235 & 231 & 11 & 1 & 0 & 1  & 2 \\ 
QSO\_RADIO\_AAL & 26 & 140 & 140 & 0 & 0 & 0 & 0  & 0 \\ 
QSO\_RADIO\_IAL & 27 & 74 & 73 & 1 & 0 & 0 & 1  & 0 \\ 
QSO\_NOAALS & 28 & 66 & 66 & 1 & 0 & 0 & 0  & 0 \\ 
QSO\_GRI & 29 & 1948 & 704 & 675 & 517 & 391 & 147  & 189 \\ 
QSO\_HIZ & 30 & 475 & 1 & 1 & 394 & 5 & 16  & 59 \\ 
QSO\_RIZ & 31 & 1344 & 87 & 83 & 1002 & 153 & 69  & 33 \\ 
BLAZGRFLAT & 50 & 102 & 59 & 8 & 10 & 7 & 22  & 4 \\ 
BLAZGRQSO & 51 & 119 & 77 & 19 & 15 & 3 & 24  & 0 \\ 
BLAZGX & 52 & 16 & 4 & 0 & 10 & 1 & 1  & 0 \\ 
BLAZGXQSO & 53 & 55 & 51 & 2 & 0 & 3 & 1  & 0 \\ 
BLAZGXR & 54 & 166 & 56 & 4 & 17 & 36 & 57  & 0 \\ 
BLAZXR & 55 & 0 & 0 & 0 & 0 & 0 & 0  & 0 \\ 
CXOBRIGHT & 58 & 188 & 130 & 5 & 3 & 50 & 5  & 0 \\ 
CXORED & 59 & 29 & 8 & 2 & 6 & 9 & 5  & 1 \\ 
\hline
ANCILLARY\_TARGET2  & & & & & & & & \\ 
\hline
HIZQSO82 & 0 & 62 & 2 & 2 & 55 & 1 & 0  & 4 \\ 
HIZQSOIR & 1 & 109 & 2 & 1 & 101 & 0 & 3  & 3 \\ 
KQSO\_BOSS & 2 & 181 & 86 & 40 & 87 & 4 & 3  & 1 \\ 
QSO\_VAR & 3 & 1394 & 877 & 313 & 429 & 85 & 0  & 3 \\ 
QSO\_VAR\_FPG & 4 & 607 & 589 & 306 & 5 & 3 & 6  & 4 \\ 
RADIO\_2LOBE\_QSO & 5 & 1133 & 590 & 63 & 345 & 113 & 62  & 23 \\ 
QSO\_SUPPZ & 7 & 4933 & 4910 & 24 & 2 & 0 & 6  & 15 \\ 
QSO\_VAR\_SDSS & 8 & 24361 & 9043 & 4223 & 13810 & 367 & 702  & 439 \\ 
QSO\_WISE\_SUPP & 9 & 5474 & 3262 & 1107 & 1726 & 329 & 134  & 23 \\ 
QSO\_WISE\_FULL\_SKY & 10 & 26966 & 25881 & 16651 & 309 & 450 & 201  & 125 \\ 
DISKEMITTER\_REPEAT & 13 & 92 & 92 & 0 & 0 & 0 & 0  & 0 \\ 
WISE\_BOSS\_QSO & 14 & 20898 & 19333 & 3867 & 395 & 933 & 115  & 122 \\ 
QSO\_XD\_KDE\_PAIR & 15 & 628 & 363 & 23 & 232 & 21 & 5  & 7 \\ 
TDSS\_PILOT & 24 & 859 & 243 & 63 & 598 & 6 & 3  & 9 \\ 
SPIDERS\_PILOT & 25 & 363 & 250 & 12 & 7 & 62 & 25  & 19 \\ 
TDSS\_SPIDERS\_PILOT & 26 & 107 & 107 & 6 & 0 & 0 & 0  & 0 \\ 
QSO\_VAR\_LF & 27 & 2401 & 1595 & 187 & 396 & 302 & 70  & 38 \\ 
QSO\_EBOSS\_W3\_ADM & 31 & 3517 & 799 & 218 & 1931 & 637 & 107  & 43 \\ 
XMM\_PRIME & 32 & 2422 & 1562 & 220 & 68 & 477 & 202  & 113 \\ 
XMM\_SECOND & 33 & 648 & 209 & 63 & 47 & 288 & 59  & 45 \\ 
SEQUELS\_TARGET & 53 & 36263 & 20333 & 3410 & 10212 & 4477 & 702  & 539 \\ 
RM\_TILE1 & 54 & 227 & 218 & 21 & 1 & 1 & 0  & 7 \\ 
RM\_TILE2 & 55 & 615 & 615 & 172 & 0 & 0 & 0  & 0 \\ 
QSO\_DEEP & 56 & 2484 & 363 & 97 & 420 & 321 & 365  & 1015 \\ 
\hline
EBOSS\_TARGET0  & & & & & & & & \\ 
\hline
QSO\_EBOSS\_CORE & 10 & 19461 & 16321 & 1208 & 679 & 1629 & 488  & 344 \\ 
QSO\_PTF & 11 & 13232 & 7485 & 935 & 3602 & 1776 & 175  & 194 \\ 
QSO\_REOBS & 12 & 1368 & 1368 & 1365 & 0 & 0 & 0  & 0 \\ 
QSO\_EBOSS\_KDE & 13 & 11843 & 9958 & 274 & 637 & 1031 & 86  & 131 \\ 
QSO\_EBOSS\_FIRST & 14 & 293 & 225 & 30 & 14 & 20 & 27  & 7 \\ 
QSO\_BAD\_BOSS & 15 & 59 & 23 & 7 & 2 & 1 & 32  & 1 \\ 
QSO\_BOSS\_TARGET & 16 & 59 & 23 & 7 & 2 & 1 & 32  & 1 \\ 
QSO\_SDSS\_TARGET & 17 & 0 & 0 & 0 & 0 & 0 & 0  & 0 \\ 
QSO\_KNOWN & 18 & 0 & 0 & 0 & 0 & 0 & 0  & 0 \\ 
SPIDERS\_RASS\_AGN & 20 & 162 & 49 & 1 & 52 & 56 & 1  & 4 \\ 
SPIDERS\_ERASS\_AGN & 22 & 0 & 0 & 0 & 0 & 0 & 0  & 0 \\ 
TDSS\_A & 30 & 9418 & 4086 & 377 & 4906 & 290 & 41  & 95 \\ 
TDSS\_FES\_DE & 31 & 42 & 42 & 0 & 0 & 0 & 0  & 0 \\ 
TDSS\_FES\_NQHISN & 33 & 74 & 73 & 0 & 0 & 0 & 1  & 0 \\ 
TDSS\_FES\_MGII & 34 & 1 & 1 & 0 & 0 & 0 & 0  & 0 \\ 
TDSS\_FES\_VARBAL & 35 & 62 & 62 & 18 & 0 & 0 & 0  & 0 \\ 
SEQUELS\_PTF\_VARIABLE & 40 & 701 & 32 & 7 & 550 & 112 & 2  & 5 \\                   
\end{longtable}
}

\newpage
 \section{Description of the superset of DR12Q}
  \label{ap:Superset_format}

\begin{table*}[h]
\centering                        
\begin{tabular}{c l c l}        
\hline\hline
Column & Name & Format & Description\\
\hline\hline
1    & SDSS\_NAME                      &  STRING      & SDSS-DR12 designation  hhmmss.ss+ddmmss.s (J2000)\\
2    & RA                              &  DOUBLE      & Right Ascension in decimal degrees (J2000)\\
3    & DEC                             & DOUBLE       & Declination in decimal degrees (J2000)\\
4    & THING\_ID                       &  INT32       & Thing\_ID \\
5    & PLATE                           & INT32        & Spectroscopic Plate number \\
6    & MJD                             & INT32        & Spectroscopic MJD \\
7    & FIBERID                         & INT32        & Spectroscopic Fiber number \\
\hline
8    & Z\_VI                           &  DOUBLE      & Redshift from visual inspection \\
9    & Z\_PIPE                         &  DOUBLE      & Redshift from BOSS pipeline \\
10   & ERR\_ZPIPE                      &  DOUBLE      & Error on BOSS pipeline redshift \\
11   & ZWARNING                        & INT32        & ZWARNING flag  \\
12   & CLASS\_PERSON                   & INT32        & Classification from the visual inspection \\
13   & Z\_CONF\_PERSON                 & INT32        & Redshift confidence from visual inspection \\
\hline
14   & SDSS\_MORPHO                    &   INT32      & SDSS morphology flag 0 = point source 1 = extended \\
15   & BOSS\_TARGET1                   &  INT64       & BOSS target flag for main survey  \\
16   & ANCILLARY\_TARGET1              &   INT64      & BOSS target flag for ancillary programs \\
17   & ANCILLARY\_TARGET2              &   INT64      & BOSS target flag for ancillary programs  \\
18   & EBOSS\_TARGET0                  &   INT64      & Target selection flag for eBOSS pilot survey \\
\hline
19   & PSFFLUX                         & FLOAT[5]     & flux in the $u$,$g$,$r$,$i$ and $z$-bands (not corrected for Galactic extinction)\\
20   & IVAR\_PSFLUX                    & FLOAT[5]     & inverse variance of $u$,$g$,$r$,$i$ and $z$ fluxes \\
21   & PSFMAG                          & FLOAT[5]     & PSF magnitudes in $u$,$g$,$r$,$i$ and $z$-bands (not corrected for Galactic extinction)\\
22   & ERR\_PSFMAG                     & FLOAT[5]     & error in $u$,$g$,$r$,$i$ and $z$ PSF magnitudes\\
23   & GAL\_EXT                        & FLOAT[5]     & Galactic extinction in the 5 SDSS bands \citep[from ][]{schlegel1998} \\
\hline
\hline
\end{tabular}
\caption{Description of the file that contains the superset from which we derive the DR12Q catalog. This file contains the result from the visual inspection as described in \Tab{VI_PIPE}.}
\label{t:superset}
\end{table*}

\end{appendix}

\end{document}